\documentclass[10pt,twoside,a4paper]{amsart}
\usepackage{amsmath}
\usepackage{amsfonts}
\usepackage{amssymb}
\usepackage{amsthm}
\usepackage{newlfont}
\usepackage{graphicx}
\usepackage{amscd}

\theoremstyle{plain}
\newtheorem{Th}{Theorem}[section]
\newtheorem{Cor}[Th]{Corollary}
\newtheorem{Lem}[Th]{Lemma}
\newtheorem{Prop}[Th]{Proposition}
\theoremstyle{definition}

\theoremstyle{remark}
\newtheorem*{Rem}{Remark}%[section]
\numberwithin{equation}{section}
\newcommand{\EE}{{\mathbb E}}

\newcommand{\FF}{{\mathbb F}}
\newcommand{\II}{{\mathbb I}}
\newcommand{\ZZ}{{\mathbb Z}}
\newcommand{\VV}{{\mathbb V}}
\newcommand{\RR}{{\mathbb R}}
\newcommand{\UU}{{\mathbb U}}
\newcommand{\WW}{{\mathbb W}}

\newcommand{\cD}{{\mathcal D}}

\newcommand{\cL}{{\mathcal L}}

\newcommand{\ba}{\boldsymbol{a}}

\newcommand{\bY}{{\boldsymbol Y}}

\newcommand{\bOm}{\boldsymbol{\Omega}}
\newcommand{\bPsi}{\boldsymbol{\Psi}}
\newcommand{\bpsi}{\boldsymbol{\psi}}
\newcommand{\bphi}{\boldsymbol{\phi}}
\newcommand{\btheta}{\boldsymbol{\theta}}
\newcommand{\bTheta}{\boldsymbol{\Theta}}

\newcommand{\bom}{\boldsymbol{\omega}}

\newcommand{\D}{{\Delta}}

\newcommand{\I}{_{(1)}}
\newcommand{\J}{_{(2)}}

\newcommand{\IMJ}{_{(-12)}}
\newcommand{\IJM}{_{(1-2)}}

\newcommand{\ii}{_{(1)}}
\newcommand{\jj}{_{(2)}}
\newcommand{\ijij}{_{(12)}}
\newcommand{\iii}{_{(11)}}
\newcommand{\jjj}{_{(22)}}
\newcommand{\iim}{_{(-1)}}
\newcommand{\jjm}{_{(-2)}}
\newcommand{\iimim}{_{(-1-1)}}
\newcommand{\jjmjm}{_{(-2-2)}}
\newcommand{\jimjm}{_{(-1-2)}}
\newcommand{\jimj}{_{(-12)}}
\newcommand{\jijm}{_{(1-2)}}

\newcommand{\Di}{\Delta_1}
\newcommand{\Dj}{\Delta_2} 
\newcommand{\Dim}{\Delta_{-1}}
\newcommand{\Djm}{\Delta_{-2}}

\newcommand{\p}{q}
\newcommand{\vt}{\vartheta}
\def\mref#1{(\ref{#1})}

\input{epsf.sty}
\usepackage{epsfig}
\usepackage{color}
\usepackage{pdflscape}
\begin{document}

\title[Darboux transformations]
{Darboux transformations for linear operators\\ on two dimensional 
regular lattices}

\author{Adam Doliwa}

\address{Adam Doliwa, Wydzia{\l} Matematyki i Informatyki,
Uniwersytet Warmi\'{n}sko-Mazurski w Olsztynie,
ul.~\.{Z}o{\l}\-nierska~14, 10-561 Olsztyn, Poland}

\email{doliwa@matman.uwm.edu.pl}
\author{Maciej Nieszporski}

\address{Maciej Nieszporski, Katedra Metod Matematycznych Fizyki, Uniwersytet
Warszawski, ul.~Ho\.{z}a~74, 00-682 Warszawa, Poland}

\email{maciejun@fuw.edu.pl}

\date{}
\keywords{integrable discrete systems; difference operators;
Darboux transformations}
\subjclass[2000]{37K10, 37K35, 37K60, 39A70}

\begin{abstract}
Darboux transformations for linear operators on regular two 
dimensional lattices are reviewed. The six point scheme is considered as the
master linear problem, whose various specifications, reductions, and their
sublattice combinations lead to other linear operators together with the
corresponding
Darboux transformations. The second part of the review deals with
multidimensional aspects of (basic
reductions of) the four point scheme, as well as
the three point scheme.
\end{abstract}
\maketitle

%\tableofcontents

\section{Introduction}
Darboux transformations are a well known tool in the theory of integrable
systems \cite{ms,RogersSchief,GuHuZhou}. The classical Darboux
transformation \cite{DarbouxIV} deals with a Sturm--Liouville problem (the one
dimensional stationary Schr\"{o}dinger equation) generating, at the same time,
new potentials and new wave functions from given ones thus providing
solutions to the Korteweg--de Vries hierarchy of $1+1$ dimensional integrable
systems. However, as it is clearly stated in \cite{DarbouxIV}, it was an 
earlier
work of Moutard \cite{Moutard} which inspired Darboux. In that work one can find
the proper transformation that applies to $2+1$ dimensional integrable systems.
Note that the initial area of applications of the Darboux transformations, which
preceded the theory of integrable systems, was the theory of conjugate
and asymptotic nets where the large body of results on Darboux
transformations was formulated 
\cite{Jonas,DarbouxIV,Eisenhart-TS,Tzitzeica,Bianchi,Lane,Finikov}.

Most of the techniques that allow us to find solutions 
of integrable non-linear differential
equations have been successfully applied to difference equations. 
These include mutually interrelated methods such as
bilinearization method \cite{Hirota} and the Sato approach \cite{DJM-I,DJM},
 direct linearization method \cite{FQC,FWN-Capel}, inverse scattering method
 \cite{Abl-Lad}, the nonlocal $\bar{\partial}$ dressing
 method \cite{BoKo}, the algebro-geometric techniques
 \cite{Krichever-4p}.

Also the method of the Darboux transformation has been successfully applied to
the discrete integrable systems. The
present paper aims to review application of 
the Darboux transformation technique for the equations that can be 
regarded as
discretizations of second order linear differential equations in two dimensions 
\eqref{2Dgeneral},
their distinguished subclasses and
systems of such equations.

While presenting the results we are trying to keep the relation with 
continuous case.
However, we are aware of weak points of this way of exposition.
The theory of discrete integrable systems \cite{Suris,DIS}
is reacher but also, in a sense,
simpler then the corresponding theory of integrable partial differential
equations. In the course of a limiting procedure, which gives differential 
systems
from the discrete ones, various symmetries and relations between different
discrete systems are lost. The classical example is provided (see, for example
\cite{BoKo-KP}) by the hierarchy of the Kadomsev--Petviashvilii (KP)
equations, which
 can be obtained from a single Hirota--Miwa equation --- 
the opposite way, from
differential to discrete, involves all
equations of the hierarchy \cite{Miwa}.

The structure of the paper is as follows. 
In section \ref{dodane} we expose main ideas in the theory of Darboux transformations in multidimension.
In Section \ref{sec:6p} 
we present the
construction of the Darboux transformation for the discrete second 
order linear problem --- the 6-point scheme \eqref{6} --- which can be 
considered as a discretization of 
the general second order linear partial differential equation in two variables
\eqref{2Dgeneral}. Then we discuss various specifications and reductions of 
the 6-point scheme. 
We separately present in Section \ref{sec:s-a} the Darboux transformations for 
discrete self-adjoint two dimensional linear
systems on the square, triangular and the honeycomb lattices, and their relation
to the discrete Moutard transformation. 
Section \ref{sec:4p} is devoted to detailed presentation of the Darboux 
transformations for systems of the 4-point linear problems, their various
specifications and the
corresponding permutability theorems. Section \ref{sec:4p-red} is dedicated  to
reductions of the fundamental transformation compatible with additional
restrictions on the form of the four point scheme. 
Finally, in Section \ref{3p-sch} we review the
Darboux transformations for the 3-point linear problem, and the corresponding
celebrated Hirota's discrete KP nonlinear system.

%%%%%%%%%%%%%%%%%%%%%%%%%%%%%%%%%%%%%%%%

%%%%%%%%%%%%%%%%%%%%%%%%%%%%%%%%%%%%%%%%%
\section{Jonas fundamental transformation and its basic reductions}
%%%%%%%%%%%%%%%%%%%%%%%%%%%%%%%%%%%%%%%%%%
\label{dodane}
The main idea included in the paper has its origin in Jonas paper \cite{Jonas} where fundamental transformation for conjugate nets has been presented. Neglecting the geometrical context of Jonas paper we only would like to say that
the fundamental transformation acts on solutions of compatible system of second order linear differential equations
on function $\psi: {\mathbb R}^n \to {\mathbb R}^m$ 
\[\psi,_{x^ix^j}=A^{ij} \psi,_{x^i}+A^{ji} \psi,_{x^j}, \qquad i,j=1,...,n  \qquad i \ne j\]
where coefficients $A^{ij}$ and $A^{ji}$ are functions of $(x^i)$, and for $n>2$ they have to obey a nonlinear differential equation (compatibility conditions); we use notation that subscript preceded by coma denotes partial differentiation with respect to indicated variables.

Every fundamental transformation $\psi \mapsto \tilde{\psi}$ can be presented as composition of  suitably chosen
\begin{enumerate}
\item radial transformation  $\psi \mapsto \psi_r$
\[\psi_r=\theta \psi\]
\item Combescure transformation $\psi_r \mapsto \tilde{\psi_r}$
\[{\tilde{\psi_r}},_{x^i}=\alpha_i {\psi_r},_{x^i}\]
\item radial transformation $\tilde{\psi_r} \mapsto \tilde{\psi}$
\[\tilde{\psi}=\gamma \tilde{\psi_r}\]
\end{enumerate}
We refer interested reader to Eisenhart book \cite{Eisenhart-TS} for further details.
As the practice shows the fundamental transformation is indeed fundamental - in the sense that most (if not all) of the  Darboux transformations are reductions of the fundamental transformation.

Here we follow this basic idea discovered almost one hundred years ago. We start from presenting
two dimensional difference operators $L$
(and corresponding difference equations $L\Psi=0$, where $\Psi$ is function of discrete variables $m$ and $n$) together with the transformation $(\Psi,L) \mapsto(\tilde{\Psi},\tilde{L})$
which is composition of
\begin{enumerate}
\item gauge transformation $(\Psi,L)\mapsto (\Psi_r,L_r)$ (see subsection \ref{GE})
\[ L_r =\Phi \circ L \circ \Theta, \qquad \Psi_r=\frac{\Psi}{\Theta}\]
where $\Theta$ is a solution of $L\Theta=0$ whereas $\Phi$ is a solution of $L^{\dagger}\Phi=0$ where $L^{\dagger}$
denotes operator formally adjoint to operator $L$. We emphasize that this particular choice of functions $\Theta$
and $\Phi$ is essential, for it guarantees existence of function $\tilde{\Psi}_r$ in the next transformation
\item transformations $(\Psi_r, L_r)\mapsto(\tilde{\Psi}_r,\tilde{L}_r)$ that take either  the form (details are given in the text below)
\begin{equation}
\label{f1}
\begin{array}{l}
\left[ 
\begin{matrix}
\Di \tilde{\Psi}_r  \cr \Dj \tilde{\Psi}_r 
\end{matrix}
\right] = 
 \left[ 
\begin{matrix}
 \alpha & \beta\cr
\gamma& \delta
\end{matrix}
\right] 
\left[ 
\begin{matrix}
\Di \Psi_r \cr  \Dj \Psi_r 
\end{matrix}
\right]
\end{array}
\end{equation}
or  the form
\begin{equation}
\label{f2}
\begin{array}{l}
\left[ 
\begin{matrix}
\Di \tilde{\Psi}_r  \cr \Dj \tilde{\Psi}_r 
\end{matrix}
\right] = 
 \left[ 
\begin{matrix}
 \alpha & \beta\cr
\gamma& \delta
\end{matrix}
\right] 
\left[ 
\begin{matrix}
\Dim \Psi_r \cr  \Djm \Psi_r 
\end{matrix}
\right]
\end{array}
\end{equation}
where $\Di$, $\Dj$ denote  forward difference operators $\Di f(m,n):= f(m+1,n)-f(m,n)$, $\Dj f(m,n):= f(m,n+1)-f(m,n)$,
whereas
$\Dim$, $\Djm$ denote  backward difference operators $\Dim f(m,n):= f(m-1,n)-f(m,n)$, $\Djm f(m,n):= f(m,n-1)-f(m,n)$.
\item gauge transformation $(\Psi_r,L_r)\mapsto (\tilde{\Psi}_r,\tilde{L}_r)$ 
\[ L_r =R \circ L \circ S, \qquad \Psi_r=\frac{\Psi}{S}\]
where in general $R$ and $S$ are arbitrary functions (which should be specified when reductions or specifications of the transformations  are considered).
\end{enumerate}

\begin{landscape}
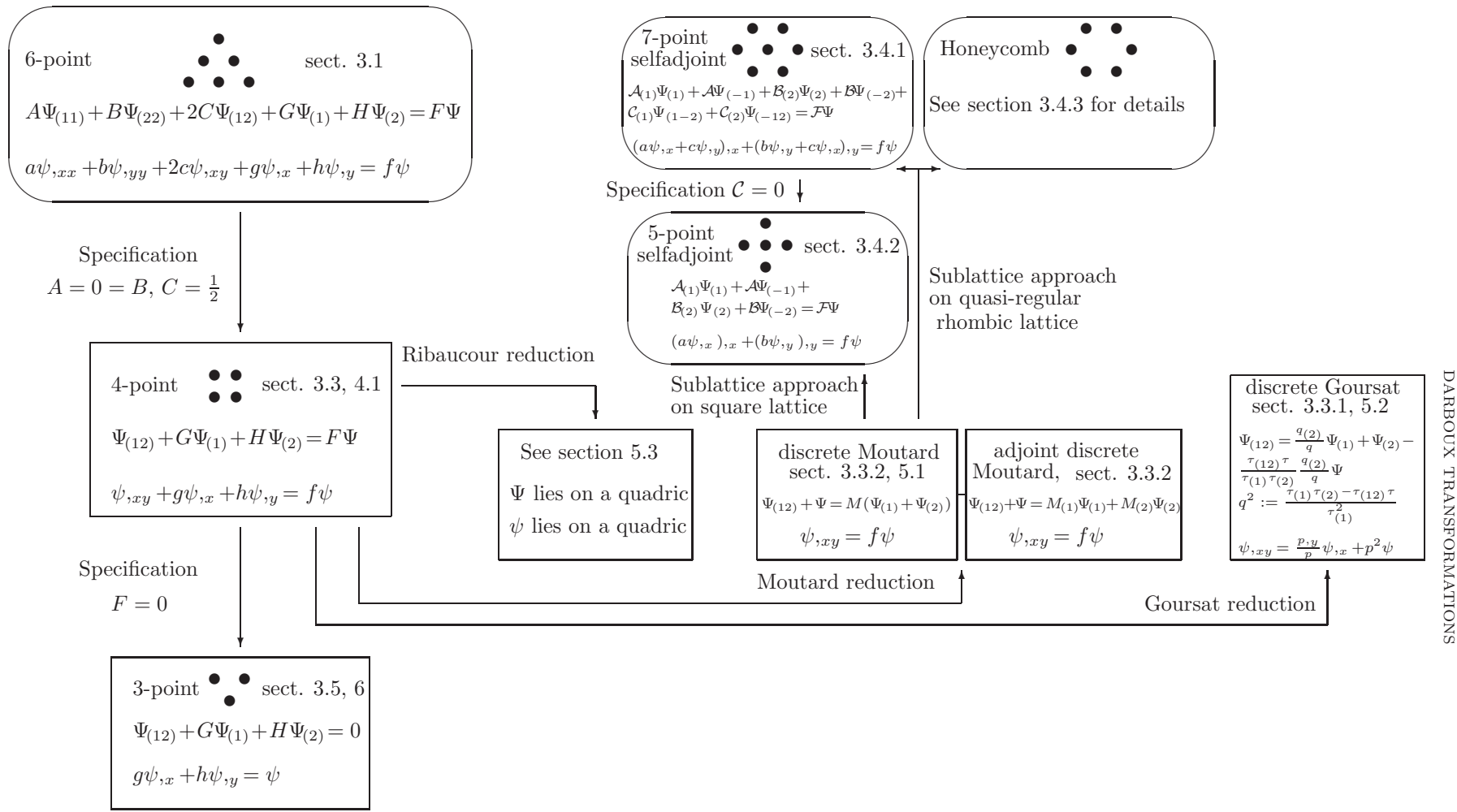
\begin{figure}%[!ht]
%\centering \fboxsep=10mm \fbox{
\begin{picture}(620,400)
%\put(70,410){\makebox(40,10)[l]{AKP}}
%\put(300,410){\makebox(40,10)[l]{BKP}}
%\put(580,410){\makebox(40,10)[l]{CKP}}

%%%%%%%%%%%%%%%%%% General Case
\put(90,360){\oval(215,90)}
\put(-10,375){\makebox(40,10)[l]{6-point}}
\put(80,390){\circle*{5}}
\put(73,380){\circle*{5}}\put(87,380){\circle*{5}} 
\put(66,370){\circle*{5}}\put(80,370){\circle*{5}} \put(94,370){\circle*{5}}
\put(120,375){\makebox(40,10)[l]{sect. \ref{Gc}}}
\put(-10,325){\makebox(50,10)[l]{$a \psi,_{xx}+b \psi,_{yy}+2 c \psi,_{xy} + g \psi,_x+ h \psi,_y=f \psi$}}
\put(-10,350){\makebox(50,10)[l]{$A \Psi \! \iii \! + \! B \Psi \! \jjj \! + \!2 C \Psi \! \ijij \!+\! G \Psi \! \ii  \!+\! H \Psi \! \jj \! =  \! F \Psi$}}

\put(90,310){\vector(0,-1){55}}
\put(15,285){\makebox(50,10)[l]{Specification}} 
\put(0,270){\makebox(50,10)[l]{$A=0=B$, $C=\frac{1}{2}$ }}

\put(20,250){\line(0,-1){80}}
\put(160,250){\line(0,-1){80}}
\put(20,250){\line(1,0){140}}
\put(20,170){\line(1,0){140}}
\put(30,225){\makebox(40,10)[l]{4-point}}
\put(78,235){\circle*{5}}\put(88,235){\circle*{5}} 
\put(78,225){\circle*{5}}\put(88,225){\circle*{5}}
\put(100,225){\makebox(40,10)[l]{sect. \ref{sec:4p-6p}, \ref{sec:fund}}}
\put(30,200){\makebox(50,10)[l]{$\Psi \! \ijij \!+\! G \Psi \! \ii  \!+\! H \Psi \! \jj \! =  \! F \Psi$}}
\put(30,175){\makebox(50,10)[l]{$\psi,_{xy} + g \psi,_x+ h \psi,_y=f \psi$}}

\put(90,165){\vector(0,-1){55}}
\put(15,140){\makebox(50,10)[l]{Specification}} 
\put(30,125){\makebox(50,10)[l]{$F=0$}}

\put(30,105){\line(0,-1){70}}
\put(150,105){\line(0,-1){70}}
\put(30,105){\line(1,0){120}}
\put(30,35){\line(1,0){120}}
\put(40,85){\makebox(40,10)[l]{3-point}}
\put(78,95){\circle*{5}}\put(92,95){\circle*{5}} 
            \put(85,85){\circle*{5}}
\put(100,85){\makebox(40,10)[l]{sect. \ref{3MN}, \ref{3p-sch}}}
\put(40,65){\makebox(50,10)[l]{$\Psi \! \ijij \!+\! G \Psi \! \ii  \!+\! H \Psi \! \jj \! = 0$ } }
\put(40,45){\makebox(50,10)[l]{$ g \psi,_x+ h \psi,_y= \psi$}}

%%%%%%%%%%%%%%%%%%%%%%%%%%%%%%%% BKP
%Ribaucour
\put(165,240){\makebox(40,10)[l]{Ribaucour reduction}}
\put(165,230){\line(1,0){90}}
\put(255,230){\vector(0,-1){15}}

\put(210,210){\line(0,-1){60}}
\put(300,210){\line(0,-1){60}}
\put(210,210){\line(1,0){90}}
\put(210,150){\line(1,0){90}}
\put(220,195){\makebox(40,10)[l]{See section \ref{Qr}}}
\put(215,175){\makebox(40,10)[l]{$\Psi$ lies on a quadric}}
\put(215,160){\makebox(40,10)[l]{$\psi$ lies on a quadric}}

\put(145,130){\line(0,1){35}}
\put(425,130){\line(-1,0){280}}
\put(425,130){\vector(0,1){15}}
\put(330,135){\makebox(50,10)[l]{Moutard reduction}}

%Moutard
\put(330,210){\line(0,-1){60}}
\put(423,210){\line(0,-1){60}}
\put(330,210){\line(1,0){93}}
\put(330,150){\line(1,0){93}}
\put(340,195){\makebox(50,10)[l]{discrete Moutard}}
\put(345,185){\makebox(40,10)[l]{sect. \ref{Mr}, \ref{bkp}}}
%\put(332,185){\makebox(50,10)[l]{equation}}
\put(332,170){\scriptsize \makebox(50,10)[l]{$\Psi \! \ijij \!+\! \Psi \!= \! M(\Psi \! \ii  \!+\!  \Psi \! \jj) $}}
\put(350,155){\makebox(50,10)[l]{$\psi,_{xy} =f \psi$}}

\put(423,180){\line(1,0){4}}
%adjoint Moutard
\put(427,210){\line(0,-1){60}}
\put(527,210){\line(0,-1){60}}
\put(427,210){\line(1,0){100}}
\put(427,150){\line(1,0){100}}
\put(440,195){\makebox(50,10)[l]{adjoint discrete}}
\put(478,185){\makebox(50,10)[l]{sect. \ref{Mr}}}
\put(430,185){\makebox(50,10)[l]{Moutard,}}
\put(428,170){\scriptsize \makebox(50,10)[l]{$\Psi \! \ijij \!\!+\!\! \Psi \!= \! M \!\ii\! \Psi \! \ii \! \!+ \! M  \!\jj \!  \Psi \! \jj $}}
\put(447,155){\makebox(50,10)[l]{$\psi,_{xy} =f \psi$}}

% 7 point
\put(335,365){\oval(135,70)}
\put(273,385){ \makebox(60,10)[l]{7-point}}
\put(268,375){ \makebox(60,10)[l]{selfadjoint}}
\put(328,395){\circle*{5}}\put(342,395){\circle*{5}} 
\put(321,385){\circle*{5}}\put(335,385){\circle*{5}} \put(349,385){\circle*{5}}
\put(328,375){\circle*{5}}\put(342,375){\circle*{5}}
\put(355,380){\makebox(40,10)[l]{sect. \ref{7ps}}}
\put(270,360){\scriptsize \makebox(100,10)[l]{${\mathcal A}\! \ii\! \Psi\! \ii\! +\! {\mathcal A}\!  \Psi\iim \! +\! 
   {\mathcal B}\! \jj\! \Psi\! \jj\! +\!{\mathcal B} \! \Psi\jjm \! 
   +$}}
\put(270,350){\scriptsize \makebox(100,10)[l]{$  {\mathcal C}\! \ii \! \Psi\jijm\! +\! {\mathcal C}\! \jj \! \Psi\! \jimj \! =\! {\mathcal F} \! \Psi$}}

\put(272,335){\scriptsize \makebox(100,10)[l]{$ (a \psi,_x\!+c \psi,_y\!\!),_x \!+(b \psi,_y\! +c\psi,_x\!\!),_y\!=\!f \psi$}}

%heksagonal
\put(475,365){\oval(135,70)}
\put(415,380){\makebox(50,10)[l]{Honeycomb}}
\put(483,395){\circle*{5}}\put(497,395){\circle*{5}} 
\put(476,385){\circle*{5}}  \put(504,385){\circle*{5}}
\put(483,375){\circle*{5}}\put(497,375){\circle*{5}}

\put(410,355){\makebox(50,10)[l]{See section \ref{hcb} for details}}

\put(405,215){\line(0,1){115}}
\put(405,330){\vector(-1,0){10}}
\put(260,315){\makebox(50,10)[l]{Specification ${\mathcal C}=0$}}
\put(405,330){\vector(1,0){10}}
\put(350,325){\vector(0,-1){10}}
\put(410,275){\makebox(50,10)[l]{Sublattice approach}}
\put(410,265){\makebox(50,10)[l]{on quasi-regular}}
\put(410,255){\makebox(50,10)[l]{
rhombic lattice}}

\put(380,215){\vector(0,1){20}}
\put(290,225){\makebox(50,10)[l]{Sublattice approach}}
\put(290,215){\makebox(50,10)[l]{on square lattice}}

% 5 point
\put(335,275){\oval(130,70)}
\put(276,295){ \makebox(60,10)[l]{5-point}}
\put(271,285){ \makebox(60,10)[l]{selfadjoint}}
                           \put(334,305){\circle*{5}}
\put(324,295){\circle*{5}} \put(334,295){\circle*{5}}\put(344,295){\circle*{5}} 
                           \put(334,285){\circle*{5}}
\put(352,290){\makebox(40,10)[l]{sect. \ref{5r}}}
\put(290,270){\scriptsize \makebox(100,10)[l]{${\mathcal A}\!\ii\!\Psi\!\ii\!+\!{\mathcal A} \!\Psi\!\iim +$}}
\put(290,260){\scriptsize \makebox(100,10)[l]{$
   {\mathcal B}\!\jj\Psi\jj\!+\!{\mathcal B}\! \Psi\!\jjm \!
    =\!{\mathcal F}\! \Psi$}} 

\put(290,245){\scriptsize \makebox(100,10)[l]{$(a \psi,_x),_x +(b \psi,_y),_y=f \psi$}}
%%%%%%%%%%%%%%%%%%%%%% CKP

\put(125,120){\line(0,1){45}}
\put(595,120){\line(-1,0){470}}
\put(595,120){\vector(0,1){25}}
\put(510,125){\makebox(50,10)[l]{Goursat reduction}}

\put(550,235){\line(0,-1){85}}
\put(640,235){\line(0,-1){85}}
\put(550,235){\line(1,0){90}}
\put(550,150){\line(1,0){90}}
\put(557,225){\makebox(50,10)[l]{discrete Goursat}}
\put(560,215){\makebox(40,10)[l]{sect. \ref{Gr}, \ref{ckp}}}

\put(551,200){\scriptsize \makebox(100,10)[l]{
$\Psi\! \ijij \!=\!\frac{ \p \jj}{ \p } \Psi \!\ii\!+ \!\Psi \!\jj - $}}
\put(551,185){\scriptsize \makebox(100,10)[l]{
$\frac{\tau \ijij \tau}{\tau \ii \tau \jj} \frac{ \p \jj}{ \p } \Psi$}}
\put(551,170){\scriptsize \makebox(100,10)[l]{
$\p ^2:= \frac{\tau \ii \tau\jj- \tau \ijij \tau }{\tau^2 \ii}$}}
\put(551,150){\scriptsize \makebox(100,10)[l]{
$\psi,_{xy} = \frac{p,_y}{p} \psi,_x+p^2\psi$}}

%\put(165,220){\vector(2,-1){100}}
%\put(165,220){\vector(3,-1){200}}
%\put(165,220){\vector(4,-1){300}}

%\put(-10,240){\makebox(50,10)[l]{$2 C \Psi \! \ijij \!+\! G \Psi \! \ii  \!+\! H \Psi \! \jj \! =  \! F \Psi$}}
%                      \put(40,81){\circle*{5}}
%\put(0,41){\circle*{5}}\put(40,41){\circle*{5}} \put(80,41){\circle*{5}}
%                      \put(40,1){\circle*{5}}
%                      \put(43,84){\scriptsize 2}
%\put(3,44){\scriptsize -1 }\put(43,44){\scriptsize  } \put(83,44){\scriptsize 1}
%                      \put(43,4){\scriptsize -2 }
\end{picture}
%} 
\caption{Interrelations between schemes discussed in the text. 
The figure include: names of the schemes used in the text, 
position of the points the equation relates, numbers of sections where the 
DT for the scheme is discussed, the equation and its natural continuum 
limit
(extensions of DT to multidimension are known only for the equations in 
rectangle frames)}
\label{5p}
\end{figure}
\end{landscape}

%\newpage
We pay special attention to
  subclasses of operators $L$  that admit
Darboux transformations. A trivial, from the point of view of integrable systems, examples are subclasses obtained by fixing a gauge (see section \ref{GE}). 
 Very important classes of operators are obtained by
 imposing conditions on the mentioned above functions $\Theta$ and $\Phi$. 
For example
in the  case of self-adjoint 7-point scheme   we discuss in section \ref{7ps}, the relation is $\Phi=\Theta$.
We refer to such procedures as to reductions.
The further examples of reductions are given in sections \ref{Gr}, \ref{Mr} and \ref{sec:4p-red}. 
We reserve separate name ``specification'' to the cases when the operator is reduced,
whereas the transformations remain essentially unaltered (i.e. no constraint on functions $\Theta$ and $\Phi$ is necessary).
Examples of specifications are given in sections  \ref{sec:4p-6p}, 
\ref{5r} and
\ref{3MN}.
%\ref{3p-sch} \ref{5r} 

With the notion of ``specification'' another important issue appears, since continuous counterparts of the specifications presented here 
can be viewed as a choice of particular gauge and independent variables.
The interesting (because not well understood) aspects of discrete integrable systems come from the fact
there is no  theory how to change discrete independent variables so that not to destroy underlying integrable phenomena.
Sublattice approach, widely used in theoretical physics, can be regarded, to some extent, as counterpart of change of independent variables. To some extent, because the only case studied in detail from the point of view of Darboux transformations
is the case of self-adjoint equation \mref{sag}, which in the discrete case consists of
Moutard case (section \ref{Mr}), self-adjoint case (section \ref{sec:s-a}) and their mutual relations    
(c.f. \cite{DGNS-4-5,DNS-4-7}). Interrelations between results 
presented here are summarized in Figure 1.

Large part of the paper is dedicated to the specification in which
the matrix in \mref{f1} is diagonal. In this case one  can consider systems of  four point operators defined on lattices with arbitrary number of independent variables. The multidimensional lattices are
extensively discussed in sections \ref{sec:4p} and \ref{sec:4p-red}, 
where compact elegant expressions for superpositions of fundamental 
transformations are presented either.

%%%%%%%%%%%%%%%%%%%%%%%%%%%%%%%%%%%%%%%%

%%%%%%%%%%%%%%%%%%%%%%%%%%%%%%%%%%%%%%%%%

\section{Two dimensional systems}
%%%%%%%%%%%%%%%%%%%%%%%%%%%%%%%%%

\label{sec:6p}
In this section we present discretizations of equation \mref{2Dgeneral} 
and its subclasses covariant under  Darboux transformations.

%\remn{specification, gauge specification, reduction}
\subsection{General case}
\label{Gc}

Out of the schemes that can serve as a discretization of the 2D equation 
\begin{equation}
\label{2Dgeneral} 
a \psi,_{xx}+b \psi,_{yy}+2 c \psi,_{xy}
+ g \psi,_x+ h \psi,_y=f \psi
\end{equation}
the following 6-point scheme
\begin{equation}
\label{6}
A \Psi \iii  +B \Psi \jjj+2 C \Psi \ijij+G \Psi \ii +H \Psi \jj =  F \Psi
\end{equation}
deserve a special attention (coefficients $A$, $B$ etc. and dependent variable $\Psi$ are functions on ${\mathbb Z}^2$, subscript in brackets denote shift operators, $f(m,n)\ii=f(m+1,n)$, $f(m,n)\jj=f(m,n+1)$, $f(m,n)\iii=f(m+2,n)$, $f(m,n)\jjj=f(m,n+2)$ and 
$f(m,n)\ijij=f(m+1,n+1)$). The scheme admits decomposition
$[(\alpha_1 T_1+\alpha_2 T_2+\alpha_3) 
(\beta_1 T_1+\beta_2 T_2+\beta_3)+\gamma]\psi=0$ (where $T_1$ and $T_2$ are forward shift operators in first and second direction respectively) and therefore its Laplace transformations can be constructed (see
\cite{Laplace,DarbouxIV,Weiss,DCN,NovVes,Adler-l,Nieszporski-ll} and 
Section~\ref{sec:Laplace-tr} for the notion of
the Laplace transformations of multidimensional linear operators).
What more important from the point of view of this review the scheme is covariant under a fundamental Darboux transformation \cite{Nieszporski-6p} (the transformation is often referred to as binary Darboux transformation in soliton literature). Indeed,
\begin{Th} \label{th:6p-Dt}
Given a non-vanishing solution $\Theta$ of \mref{6}
\begin{equation}
\begin{array}{l}
\label{6t}
 A \Theta\iii  +B \Theta\jjj+2 C \Theta\ijij+G \Theta\ii  +H \Theta\jj =  F \Theta
\end{array}
\end{equation}
and a non-vanishing solution $\Phi$ of the equation adjoint to equation \mref{6}
\begin{equation}
\begin{array}{l}
\label{6ad}
A\iimim \Phi\iimim +B\jjmjm \Phi\jjmjm+2 C\jimjm \Phi\jimjm
+ G\iim  \Phi\iim  +H\jjm  \Phi\jjm =F \Phi
\end{array}
\end{equation}
(negative integers in brackets in  subscript denote backward shift e.g. $f(m,n)\iim=f(m-1,n)$ etc.)
the existence of auxiliary function $P$ is guaranteed
\begin{equation}
\label{P}
\begin{array}{l}
\Dim (\Phi\Theta \ijij P) = 
B\jjm  \Phi\jjm  \Theta\jj+B \Phi \Theta\jjj+
C\iim \Phi\iim \Theta\jj+C \Phi \Theta\ijij+H \Phi \Theta\jj \\
\Djm (\Phi\Theta \ijij P) =
-(A\iim  \Phi\iim  \Theta\ii+A \Phi \Theta\iii +
C\jjm \Phi\jjm \Theta\ii+C \Phi \Theta\ijij+G \Phi \Theta\ii).
\end{array}
\end{equation}
Then equation \mref{6} can be rewritten as
\begin{equation}
%\label{}
\begin{array}{l}  \Dj \left[ (C\jjm + P\jjm )\Phi\jjm \Theta\ii \Di \left(\frac{\Psi}{\Theta} \right) +
B\jjm  \Phi\jjm  \Theta\jj \Dj \left(\frac{\Psi}{\Theta} \right) \right]+\\
\Di \left[ A\iim \Phi\iim  \Theta\ii\Di \left(\frac{\Psi}{\Theta} \right) -(P\iim - C\iim )\Phi\iim \Theta\jj)
\Dj \left(\frac{\Psi}{\Theta} \right) \right]=0
\end{array}
\end{equation}
which in turn guarantees
the existence of functions  $\Psi'$ such that
\begin{equation}
\label{DT6}
\begin{array}{l}
\left[ 
\begin{matrix}
\Di (\Psi') \cr \Dj (\Psi')
\end{matrix}
\right] = 
 \left[ 
\begin{matrix}
 (C\jjm + P\jjm )\Phi\jjm \Theta\ii& B\jjm  \Phi\jjm  \Theta\jj\cr
-A\iim \Phi\iim  \Theta\ii&(P\iim - C\iim )\Phi\iim \Theta\jj
\end{matrix}
\right] 
\left[ 
\begin{matrix}
\Di \left(\frac{\Psi}{\Theta} \right)\cr  \Dj \left(\frac{\Psi}{\Theta} \right)
\end{matrix}
\right]
\end{array}
\end{equation}
Assuming that matrix in \mref{DT6}
is invertible on the whole lattice, i.e.
\newline
{\small$D:=[(P\iim - C\iim )(P\jjm \!+\! C\jjm ) + A\iim  B\jjm ] \Theta\ii \Theta\jj \Phi\iim  \Phi\jjm \ne 0$}
everywhere and
finally on introducing $\tilde{\Psi}$ via
\begin{equation}
\label{S}
\tilde{\Psi}=\frac{\Psi'}{S} 
\end{equation}
and  taking the opportunity of multiplying resulting equation by non-vanishing function $R$,
we arrive at the conclusion that the function $\tilde{\Psi}$ satisfies equation of the form \mref{6} but with new coefficients
\begin{equation}
\label{dcoef}
\begin{array}{l}
\tilde{A}=\frac{R S\iii  \Phi \Theta\iii }{D\ii} A,
\qquad
\tilde{B}=\frac{R S\jjj \Phi \Theta\jjj }{D\jj} B,
\qquad
\tilde{F}=\frac{R S \Phi \Theta }{D} F,
\\ 
\\
\tilde{C} =\frac{R S\ijij}{2}\left(
\frac{\Theta\iii \Phi\jijm(C+P)\jijm}{D\ii}+\frac{\Theta\jjj\Phi\jimj(C-P)\jimj}{D\jj}\right), 
\\
\\
{\tilde{G}}=-{R S\ii}\left( \frac{\Theta\iii \Phi\jijm(C+P)\jijm+\Theta\iii \Phi A}{D\ii}+
                                 \frac{\Theta\jj\Phi\iim (C-P)\iim +\Theta\ii\Phi\iim A\iim }{D} \right),
\\
\\
{\tilde{H}}=-{R S\jj}\left( \frac{\Theta\jjj\Phi\jimj(C-P)\jimj+\Theta\jjj\Phi B}{D\jj}
                                +\frac{\Theta\ii\Phi\jjm (C+P)\jjm +\Theta\jj\Phi\jjm B\jjm }{D}\right),
\end{array}
\end{equation}
\end{Th}

The family of maps $\Psi \mapsto \tilde{\Psi}$ given by \mref{DT6}-\mref{S} we refer to as Darboux transformations of
equation \mref{6}.

\subsection{Gauge equivalence} 
%%%%%%%%%%%%%%%%%%%%%%%%%%%%%%
%In the theory of integrable systems the basic object are classes of gauge equivalent linear equations (not the equation itself).
\label{GE}
We say that two linear operators $L$ and $L'$ are gauge equivalent
if one can find functions say $\Phi$ and $\Theta$ such that $L'=\Phi \circ L \circ \Theta$ (where $\circ$ stands for composition of operators).
The idea to consider equivalence classes of linear operators \mref{2Dgeneral} with respect to gauge  rather than single operator itself,
goes to Laplace and Darboux papers \cite{Laplace,DarbouxIV}. 
In the continuous case it reflects in the fact that one can confine himself e.g.
to equations (the so called affine gauge)
\[a \psi,_{xx}+b \psi,_{yy}+ \psi,_{xy}
+ g \psi,_x+ h \psi,_y=0\]
or  to (in this case we would like to introduce the name {\em basic gauge}) 
\[a \psi,_{xx}+b \psi,_{yy}+ 2c\psi,_{xy}
+ k,_y \psi,_x -k,_x \psi,_y=0\]
without loss of generality.

%Gauge transformation $L\mapsto L'=\Phi \circ L \circ \Theta$, is one of the ingredients of the fundamental Darboux transformation presented in the previous section (it can be regarded a composition of gauge transformation $L \mapsto L$. Moreover, 
The Darboux transformation can be viewed as transformation acting on equivalence classes (with respect to gauge) of equation \mref{6} (compare \cite{Nieszporski-6p}).
Therefore one can confine himself to particular elements of equivalence class. 
Commonly used choice is to confine oneself to the  affine gauge i.e. to equations \mref{6} that obeys
\[A + B  + 2C + G + H - F  = 0.\]
If one puts S = const in \mref{S} 
(this condition is not necessary) then the above constraint is preserved under the Darboux transformation.
One can consider further specification of the gauge
\[A + B  + 2C + G + H - F  = 0, \quad A\IMJ + B\IJM + 2C  + G \jj + H \ii - F \ijij = 0 \]              
This choice of gauge we would like to refer as to basic gauge of equation \mref{6}. Note that if the equation \mref{6} is in a basic gauge its formal adjoint is in a basic gauge too.

\subsection{Specification to 4-point scheme and its reductions}
%%%%%%%%%%%%%%%%%%%%%%%%%%%%%%%%%%%%%%%%%%%%%%%%%%%%%%%%%%%%%%
\label{sec:4p-6p}
In the continuous case due to possibility of changing independent variables one can reduce, provided  equation \mref{2Dgeneral} is hyperbolic, to canonical form
\[\psi,_{xy}+ g \psi,_x+ h \psi,_y=f \psi.\]
In the discrete case similar result can be obtained in a different way.
Taking a glance at  (\ref{dcoef}) we can notice that coefficients $A$, $B$ and $F$ transform in a very simple manner. In particular, if any of these coefficients equals  zero then its transform equals  zero too.
 Let us stress that in this case if we do so, we do not impose any constraints on transformation data, transformations remains essentially the same.
First  we shall  concentrate on the case when two out of three mentioned functions vanish. 

If we put
\begin{equation}
\label{4ps}
A=0, \quad B=0, \quad C=\frac{1}{2}
\end{equation}
then from equation \mref{dcoef}
\[
\tilde{A}=0, \quad \tilde{B}=0
\]
and one can adjust the function  $R$ so that
\[\tilde{C}=\frac{1}{2}\]
so we arrive at the 4-point scheme
\begin{equation}
\label{4pabc}
\Psi \ijij=\alpha \Psi \ii+\beta \Psi \jj +\gamma \Psi.
\end{equation}
and its fundamental transformation c.f. \cite{BoKo,MQL,MDS,TQL}.

We observe that the form of equation \mref{4pabc} is covariant under the gauge $L\mapsto \frac{1}{g \ijij} L g$
and to identify whether two equations are equivalent or not we use the invariants of the gauge
\begin{equation}
%\label{}
\kappa=\frac{\alpha\jj\beta }{\gamma\jj}, \quad n=\frac{\alpha \beta \ii}{\gamma\ii}
\end{equation}
Two equations are equivalent if their corresponding invariants $\kappa$ and $n$ are equal \cite{Nieszporski-ll}.

\subsubsection{Goursat equation} 
%%%%%%%%%%%%%%%%%%%%%%%%%%%%%%%%%%%%%%%%%%%%%%%%%%%%%%%%%%%%%%%%%%%%%%%%%%%%%%
\label{Gr}
In this subsection we discuss the discretization of class of equations
\[\psi,_{xy} = \frac{p,_x}{p} \psi,_y+p^2\psi\]
which is referred to as  Goursat equation.
The discrete counterpart of Goursat equation arose from the surveys on Egorov lattices
\cite{Schief-talk} and symmetric lattices 
 \cite{DS-sym} and can be written in the form \cite{Nieszporski-ll}
\begin{equation}
\label{dG}
\Psi \ijij =\frac{ \p \jj}{ \p } \Psi \ii+ \Psi \jj -\frac{\tau \ijij \tau}{\tau \ii \tau \jj} \frac{ \p \jj}{ \p } \Psi
\end{equation}
where functions $ \p $ and $\tau$ are related via
\begin{equation}
\p ^2= \frac{\tau \ii \tau\jj- \tau \ijij \tau }{\tau^2 \ii}.
\end{equation}
The gauge invariant characterization of the discrete Goursat equation is  either
\[n\jj ^2=\kappa \kappa\ijij \frac{(1+\kappa \ii)(1+\kappa \jj)}{(1+\kappa )(1+\kappa \ijij)}  \]
or
\[\kappa\ii ^2=n n\ijij \frac{(1+n \ii)(1+n \jj)}{(1+n )(1+n \ijij)}  \]

The Goursat equation can be isolated from the others 4-point schemes in the similar way that Goursat did it over hundred years ago \cite{Goursat,Nieszporski-ll} i.e. as the equation such that 
one of its Laplace transformations maps solutions of the equation to solutions of the adjoint equation.
Therefore in this case  if $\Theta$ obeys \mref{dG}
\begin{equation}
\label{dG1}
\Theta \ijij =\frac{ \p \jj}{ \p } \Theta \ii+ \Theta \jj -\frac{\tau \ijij \tau}{\tau \ii \tau \jj} \frac{ \p \jj}{ \p } \Theta
\end{equation}
then its Laplace transformation
\begin{equation}
\label{fGt}
\Phi=\frac{1}{\p ^2\jj}\frac{\tau\jj}{\tau \ijij} \Di \Theta \jj
\end{equation}
is a solution of equation adjoint to equation \mref{dG}  \cite{Nieszporski-ll}. 
Equation \mref{fGt} is the constraint we impose on transformation data (i.e. functions $\Theta$ and $\Phi$) in the fundamental transformation
\mref{DT6}, \mref{dcoef} (we recall we have already put $A$ and $B$ equal to zero).
In addition if $\Theta$ obeys \mref{dG1} then
\begin{equation}
%\label{}
\Di\left(\frac{\tau}{\tau \jj}\Theta^2 \right)=\Dj\left[\frac{1}{\p ^2}\frac{\tau}{\tau \ii}(\Di\Theta)^2 \right]
\end{equation}
and as a result there exists function $\vt$ such that
\begin{equation}
%\label{}
\Di \vt=\frac{1}{\p ^2}\frac{\tau}{\tau \ii}(\Di\Theta)^2, \quad\Dj\vt=\frac{\tau}{\tau \jj}\Theta^2
\end{equation}
Now it can be shown that one can put
\begin{equation}
%\label{}
\Phi \Theta \ijij (P-\frac{1}{2})=-\vt \ijij , \quad \Phi \Theta \ijij (P+\frac{1}{2})= 
\left[\sqrt{\frac{\tau \jj}{\tau \ii}}\frac{\sqrt{\Di \vt \Dj \vt}}{q} \right] \jj -\vt\jj
\end{equation}
and the transformation \mref{DT6} takes form
\begin{equation}
\label{GDT}
\begin{array}{l}
\left[ 
\begin{matrix}
\Di (\sqrt{\frac{{\tau }}{{\tau\jj}}} \frac{\vt}{\sqrt{\Dj\vt}}\tilde{\Psi}) \cr 
\Dj (\sqrt{\frac{{\tau }}{{\tau \jj}}} \frac{\vt}{\sqrt{\Dj\vt}}\tilde{\Psi})
\end{matrix}
\right] = 
 \left[ 
\begin{matrix}
  \sqrt{\frac{\tau \jj}{\tau \ii}}\frac{\sqrt{\Di \vt \Dj \vt}}{q} -\vt & 0 \cr
0 & - \vt \jj
\end{matrix}
\right] 
\left[ 
\begin{matrix}
\Di \left(\sqrt{\frac{{\tau}}{{\tau \jj}}}  \frac{\Psi}{\sqrt{\Dj\vt}}\right)\cr  
\Dj \left(\sqrt{\frac{{\tau}}{{\tau \jj}}}  \frac{\Psi}{\sqrt{\Dj\vt}} \right)
\end{matrix}
\right]
\end{array}
\end{equation}
which is the discrete version of Goursat transformation
\cite{Goursat}. The transformation rule for the field $\tau$ is
\[\tilde\tau=\tau \vt.\]

\subsubsection{Moutard equation}
%%%%%%%%%%%%%%%%%%%%%%%%%%%%%%%%
\label{Mr}
In this subsection we discuss discretization Moutard equation and its (Moutard) transformation \cite{Moutard}
\[\psi,_{xy} = f \psi.\]
Moutard equation is self-adjoint equation
(which allowed us to impose reduction $\Phi=\Theta$ in the continuous analogue of transformation \mref{DT6}
c.f. \cite{Nieszporski-6p}).
The point is that opposite to the continuous case   there is no appropriate self-adjoint 4-point scheme 

%Let us trace the 
%Moutard reduction of Fundamental transformation to reveal its adjoint counterpart and analogue of constraint $\Phi=\Theta$ .

We do have the reduction of fundamental transformation that can be regarded as discrete counterpart of Moutard transformation.
Namely, class of equations that can be written in the form
\begin{equation}
\label{dM}
\Psi \ijij + \Psi=M(\Psi \I+ \Psi \J)
\end{equation}
we refer nowadays to as discrete Moutard equation. It appeared in the context of integrable systems in 
\cite{DJM} and then its Moutard transformations have been studied in  detail  in \cite{NiSchief}.
Gauge invariant characterization of the class of discrete Moutard equations is \cite{Nieszporski-ll}
\[n\jj n=\kappa \ii \kappa\]
Let us trace this reduction on the level of fundamental transformation.
Putting 
$2C=-F$  and $G=H=:M F$
the equations \mref{6}, \mref{6t} and \mref{6ad} take respectively form
\begin{equation}
\begin{array}{l}
\label{6M}
F (\Psi \ijij +\Psi)= G (\Psi\ii + \Psi\jj)      
\end{array}
\end{equation}

\begin{equation}
\begin{array}{l}
\label{6tM}
F(\Theta\ijij +\Theta)= G (\Theta\ii + \Theta\jj) 
\end{array}
\end{equation}

\begin{equation}
\begin{array}{l}
\label{6adM}
F\jimjm \Phi\jimjm+F \Phi= G\iim   \Phi\iim  + G\jjm   \Phi \jjm  
\end{array}
\end{equation}
The crucial observation is:
if the function $\Theta$ satisfies equation \mref{6tM} then the function
$\Phi$ given by
\begin{equation}
\label{PT}
\begin{array}{l}
\Phi=\frac{1}{F} (\Theta\ii+ \Theta\jj)
\end{array}
\end{equation}
satisfies equation \mref{6adM} \cite{Nieszporski-ll}.
If we put
\[2 P=\frac{\Theta\ii- \Theta\jj}{\Phi}\]
then equations \mref{P} will be automatically satisfied. If in addition we put in \mref{S} $S=\Theta$ then
Darboux transformation \mref{DT6} takes form (c.f. \cite{NiSchief}) 
\begin{equation}
\label{MDT}
\left[ 
\begin{matrix}
\Di (\Theta \tilde{\Psi}) \cr \Dj (\Theta \tilde{\Psi})
\end{matrix}
\right]
= \left[ 
\begin{matrix} \Theta \Theta \I & 0\cr
 0  & -\Theta \Theta \J 
\end{matrix}
\right]
\left[ 
\begin{matrix}  \Di \frac{\Psi}{\Theta}\cr \Dj \frac{\Psi}{\Theta}
\end{matrix}
\right]
\end{equation}
and it serves as transformation $\Psi \mapsto \tilde{\Psi}$ that maps solutions of equation \mref{dM}
 into solutions of another discrete Moutard equation
\begin{equation}
%\label{dM}
\tilde{\Psi}\ijij + \tilde{\Psi}=\widetilde{M}( \tilde{\Psi}\I+ \tilde{\Psi} \J), \quad 
\widetilde{M}=\frac{\Theta \I\Theta \J}{\Theta \ijij\Theta}M
%\frac{\frac{1}{\Theta \ijij} + \frac{1}{\Theta}}{\frac{1}{\Theta \I}+ \frac{1}{\Theta\J}}
\end{equation}
We recall that  $\Theta$ an arbitrary non-vanishing fixed solution of the equation \mref{dM}
\begin{equation}
\label{Ms}
 \Theta \ijij + \Theta=M(\Theta \I+ \Theta \J)
\end{equation}

Every equation that is gauge equivalent to \mref{6adM} we refer to as an adjoint discrete equation Moutard equation, its gauge invariant characterization is \cite{Nieszporski-ll}
\[\kappa \ijij \kappa \ii = n\ijij n \jj.\]
Equation \mref{6tM} can be regarded as potential version of equation \mref{6adM}

\subsection{Self-adjoint case}
\label{sec:s-a}
%%%%%%%%%%%%%%%%%%%%%%%%%%%%%%%%%%%%%%%%%%%%%%%%%%%%%%%%%%%%%%%%
In the continuous case there is direct reduction of 
the fundamental transformation 
for equation \mref{2Dgeneral} to the Moutard type transformation for 
self-adjoint equation  \cite{Nieszporski-6p}
\begin{equation}
\label{sag}
(a \psi,_x+c \psi,_y),_x +(b \psi,_y +c\psi,_x),_y=f \psi.
\end{equation}
In this Section we present
difference analogues of equation \eqref{sag} which allow for the Darboux
transformation.Opposite to the continuous case the transformation 
will not be direct reduction of the fundamental transformation for the 6-point 
scheme \mref{6}. 
The self-adjoint discrete operators studied below are however intimately 
related to discrete Moutard equation which provides the link between their Darboux
transformations and the fundamental transformation for equation \mref{2Dgeneral}.

\subsubsection{7-point self-adjoint scheme}
%%%%%%%%%%%%%%%%%%%%%%%%%%%%%%%%%%%%%%%
\label{7ps}
The following 7-point linear system
\begin{equation}
\label{7eq}
{\mathcal A}\ii\Psi\ii+{\mathcal A} \Psi\iim +
   {\mathcal B}\jj\Psi\jj+{\mathcal B} \Psi\jjm 
   +  {\mathcal C}\ii \Psi\jijm+{\mathcal C}\jj \Psi\jimj ={\mathcal F} \Psi,
\end{equation}
allows for the Darboux transformation \cite{NSD}.
\begin{Th} \label{th:tp-D}
Given scalar solution $\varTheta$ of the linear equation \eqref{7eq}, then
$\tilde\Psi$ given as solution of the following system
\begin{equation} 
\begin{pmatrix}
\Di (\tilde{\Psi}  \varTheta) \\ \Dj ( \tilde{\Psi} \varTheta) 
\end{pmatrix} =
\begin{pmatrix}
 {\mathcal C} \varTheta\iim  \varTheta\jjm &
-\varTheta\jjm  ({\mathcal B} \varTheta  +
{\mathcal C} \varTheta\iim  ) \\
\varTheta\iim  \left( {\mathcal A} \varTheta  \!+\! {\mathcal C}
 \varTheta\jjm    \right) &
- {\mathcal C} \varTheta\iim  \varTheta\jjm 
\end{pmatrix}
\begin{pmatrix}
\Dim \frac{\Psi}{\varTheta}
\\
\Djm \frac{\Psi}{\varTheta} 
 \end{pmatrix}
\end{equation}
satisfies the 7-point scheme \eqref{7eq}
with the new fields given by
\begin{equation}
\tilde{{\mathcal A}}\ii=\frac{\varTheta \varTheta\ii 
{\mathcal A}}{\varTheta\jjm   {\mathcal P}},  
\qquad
\tilde{{\mathcal B}}\jj=\frac{\varTheta \varTheta\jj {\mathcal B}}{\varTheta\iim   {\mathcal P}},
, \qquad
\tilde{{\mathcal C}}=\frac{{\mathcal C}\jimjm \varTheta\iim  \varTheta\jjm  }
{\varTheta\jimjm {\mathcal P}\jimjm}, 
\end{equation}
\begin{equation}
\tilde{{F}}=\varTheta \left(\frac{\tilde{{\mathcal A}}\ii }{\varTheta\ii}+   
\frac{\tilde{{\mathcal A}} }{\varTheta\iim }+  
\frac{\tilde{{\mathcal B}}\jj }{\varTheta\jj} + 
 \frac{\tilde{{\mathcal B}}}{\varTheta\jjm }+
 \frac{\tilde{{\mathcal C}}\ii}{\varTheta\jijm}+
\frac{\tilde{{\mathcal C}}\jj }{\varTheta\jimj}\right)
\end{equation}
where
\begin{equation}
{\mathcal P}= \varTheta {\mathcal A}{\mathcal B} +
\varTheta\iim  {\mathcal C} {\mathcal A}+\varTheta\jjm  
{\mathcal C}  {\mathcal B}.
\end{equation} 
\end{Th}
\begin{figure}
\begin{center}
\includegraphics[width=7cm]{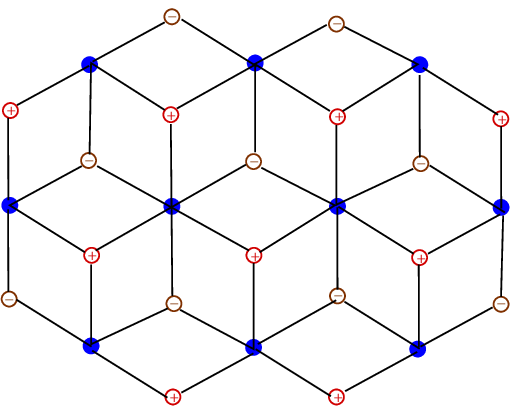} \hskip1cm
\includegraphics[width=7cm]{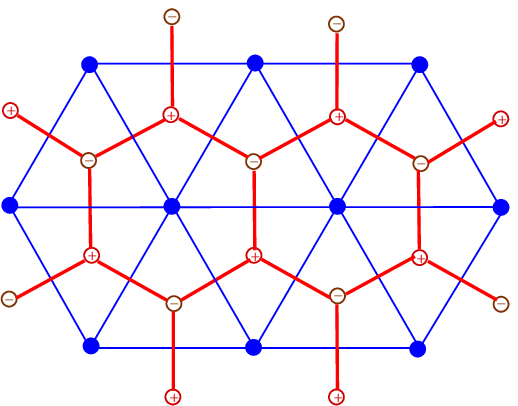}
\end{center}
\caption{The bipartite quasi-regular rhombic
tiling and the triangular and honeycomb lattices}
\label{fig:QRRT}
\end{figure}  
As it was shown in \cite{DNS-4-7} the self-adjoint 7-point scheme \eqref{7eq}
can be obtained from the system of Moutard equations imposed consistently on 
quadrilaterals of the bipartite quasi-regular rhombic tiling (see Fig.~\ref{fig:QRRT}), which is a 
particular case of the
approach considered in \cite{BobMerSur}. Then the Moutard transformations can be
also restricted to the triangular sublattice leading to Theorem \ref{th:tp-D}.

\subsubsection{Specification to  5-point scheme}
%%%%%%%%%%%%%%%%%%%%%%%%%%%%%%%%%%%%%%%%%%%%%%%%
\label{5r}
The 7-point scheme admits specification ${\mathcal C}=0$ 
(alternatively one can put ${\mathcal A}=0$ or ${\mathcal B}=0$) and 
as result we obtain specification
to 5-point self-adjoint scheme  \cite{NSD}. 
\begin{equation}
\label{5eq}
 {\mathcal A}\ii\Psi\ii+{\mathcal A} \Psi\iim +
   {\mathcal B}\jj\Psi\jj+{\mathcal B} \Psi\jjm 
    ={\mathcal F} \Psi .
\end{equation}
The self-adjoint 5-point scheme and its Darboux transformation can be also
obtained form the Moutard equation on the (bipartite)
square lattice \cite{DGNS-4-5}.

\subsubsection{The honeycomb lattice}
%%%%%%%%%%%%%%%%%%%%%%%%%
\label{hcb}
It is well known that the triangular and honeycomb grids are dual to each other
(see Fig.~\ref{fig:QRRT}). Restriction of the system of the Moutard equations on
the rhombic tiling to the honeycomb sublattice gives \cite{DNS-4-7}
the following linear system
\begin{align} \label{eq:hcl1}
\frac{1}{{\mathcal A}}(\Psi^+ - \Psi^-) + 
\frac{1}{{\mathcal B}}(\Psi^+_{(-1 2)} - \Psi^-) +
\frac{1}{{\mathcal C}_{(1 2)}}(\Psi^+_{(2)} - \Psi^-) & = 0,\\
\label{eq:hcl2}
\frac{1}{{\mathcal A}}(\Psi^- - \Psi^+) + 
\frac{1}{{\mathcal B}_{(1 -2)}}(\Psi^-_{(1 -2)} - \Psi^+) +
\frac{1}{{\mathcal C}_{(1)}}(\Psi^-_{(-2)} - \Psi^+) & = 0.
\end{align}
\begin{Rem}
Because $\Psi^-$ and $\Psi^+$ satisfy separately the
self-adjoint 7-point schemes \eqref{7eq}, but with different coefficients, then 
the linear problem \eqref{eq:hcl1}-\eqref{eq:hcl2} can be considered as a
relation between two equations \eqref{7eq}. This is the Laplace transformation
between self-adjoint 7-point schemes studied in
\cite{NovVes,NovDyn}.
\end{Rem}
The corresponding restriction of the Moutard transformation gives the Darboux
transformation for the honeycomb linear problem. 
\begin{Th} \label{th:DT-hc}
Given scalar solution $(\varTheta^+,\varTheta^-)$ of the 
 honeycomb linear system \eqref{eq:hcl1}-\eqref{eq:hcl2}
then the solution $(\tilde\Psi^+,\tilde\Psi^-)$ of the system
\begin{align} 
\label{eq:DT-hc1}
\tilde\Psi^+_{(2)} - \tilde\Psi^- & = \frac{{\mathcal C}}{{\mathcal R}}\left( 
\varTheta^-_{(-2)}\Psi^-_{(-1)} - \theta^-_{(-1)}\Psi^-_{(-2)} \right),\\
\label{eq:DT-hc2}
\tilde\Psi^+ - \tilde\Psi^- & = \frac{{\mathcal A}_{(-1)}}{{\mathcal R}}\left(
\varTheta^-_{(-1-2)}\Psi^-_{(-2)} -\varTheta^-_{(-2)}\Psi^-_{(-1-2)} \right), \\
\label{eq:DT-hc3}
\tilde\Psi^- - \tilde\Psi^+_{(-1 2)} & = \frac{{\mathcal B}_{(-2)}}{{\mathcal R}}\left(
\varTheta^-_{(-1-2)}\Psi^-_{(-1)} -\varTheta^-_{(-1)}\Psi^-_{(-1-2)} \right),
\end{align}
with ${\mathcal R}$ given by 
\begin{equation}
{\mathcal R}= {\mathcal A}_{(-1)}{\mathcal B}_{(-2)} +
{\mathcal C} {\mathcal A}_{(-1)}+ 
{\mathcal C}  {\mathcal B}_{(-2)},
\end{equation}
satisfies a new honeycomb linear system with the
coefficients 
\begin{equation} \label{eq:coeff-DT-hc}
\tilde{\mathcal A} = \frac{{\mathcal A}_{(-1)}}{{\mathcal R}} 
\varTheta^{-}_{(-2)}\varTheta^{-}_{(-1-2)}, \qquad
\tilde{\mathcal B} = \frac{{\mathcal B}_{(-2)}}{{\mathcal R}} 
\varTheta^{-}_{(-1)}\varTheta^{-}_{(-1-2)}, \qquad
\tilde{\mathcal C}_{(12)} = \frac{{\mathcal C}}{{\mathcal R}} 
\varTheta^{-}_{(-1)}\varTheta^{-}_{(-2)}.
\end{equation}
\end{Th}

\subsection{Specification to 3-point scheme}
%%%%%%%%%%%%%%%%%%%%%%%%%%%%%%%%%%%%%%%%%%%%
\label{3MN}
We end  the review of two-dimensional case with specification to the 3-point scheme 
i.e. to a discretization of
first order differential equation
\[g \psi,_x+ h \psi,_y=f\psi.\]
If we put 
\[A=0,\quad B=0, \quad F=0\]
then according to \mref{dcoef}
 \[\tilde{A}=0,\quad \tilde{B}=0, \quad \tilde{F}=0\]
It means that the fundamental transformation \mref{DT6} is also the Darboux transformation
for the 3-point scheme \cite{DJM-II,Nimmo-KP}
\[2C \Psi \ijij + G\Psi \ii +H\Psi \jj =0 .\]
This elementary scheme is the simplest one from the class considered here, but 
it deserves a special attention, because it leads to one of the most studied
integrable discrete equation \cite{Hirota}.
We confine ourselves to recalling briefly
in Section \ref{3p-sch} main results in this field.

To the end let us rewrite the 3-point scheme in the basic gauge
\begin{equation} 
(u\jj-u\ii)\Psi \ijij+(u\ii-u)\Psi\ii-(u\jj-u)\Psi \jj=0
\end{equation}
%\begin{equation} \label{eq:BKP-lin}
%\bx_{(ij)} - \bx = F_{ij} (\bx_{(i)} - \bx_{(j)}) , \quad 1\leq i< j\leq N,
%\end{equation}

%\begin{displaymath}
%\begin{array}{ll}
%\hbox{Jonas}  & \psi,_{xy}+ g \psi,_x+ h \psi,_y=f \psi \\
%\hbox{Moutard} & \psi,_{xy} = f \psi\\
%\hbox{Goursat} & \psi,_{xy} = \frac{p,_x}{p} \psi,_y+p^2\psi \\
%\hbox{Ribaucour I} &  \rr,_{xy}+w \rr,_x+z \rr,_y=0 \\
%&   \qquad \rr \hbox{ lies on a quadric}\\
%\hbox{Ribaucour II} & \rr,_{xy}+w \rr,_x+z \rr,_y=0 \\
%&|\rr|^2,_{xy}+w |\rr|^2,_x+z |\rr|^2,_y=0
%\end{array}
%\end{displaymath}

%%%%%%%%%%%%%%%%%%%%%%%%%%%%%%%%%%%%%%%%%

%%%%%%%%%%%%%%%%%%%%%%%%%%%%%%%%%%%%%%%%%

\section{The four point systems}
\label{sec:4p}
In this Section we present the Darboux transformations for the four point
scheme (the discrete Laplace equation) from the point of view of systems of such
equations, and the corresponding permutability theorems.
To keep the paper of reasonable size and in order to present the results from
a simple algebraic perspective we do not discuss important relations of the 
subject to incidence and difference geometry
\cite{DCN,MQL,Bobenko-DO,CDS,q-red,Dol-Koe,BQL,CQL,TQL,DS-sym,KingSchief1,KingSchief,
KoSchief2,KoSchiefSBKP,Schief-JNMP,Schief-II,DGNS-4-5,DNS-4-7} (see also \cite{DS-EMP,BobSur}
and earlier works \cite{Sauer2,Sauer}), application of
analytic \cite{BoKo,DMS,TQL,DMMMS,WTLS,WTS,WOGTS,DMM,DS-sym,MM-2,AD-tauQL} and 
algebro-geometric 
\cite{Krichever-4p,KWZ,AKV,AD-DQL,AD-Chicago,BD,DGNS-4-5,BQL,GrKri,CQL}
techniques of the integrable systems theory to construct large classes of
solutions of the linear systems in question and solutions of the corresponding
nonlinear discrete equations.

In order to simplify discussion of the Darboux transformations for systems of
the 4-point schemes we fix (without loss of generality \cite{MQL}) the gauge to
the affine one
\begin{equation} \label{eq:4p-affine}
\bPsi_{(ij)} = A_{ij(i)}\bPsi_{(i)}  + 
A_{ji(j)}\bPsi_{(j)}  +(1 - A_{ij(i)} - A_{ji(j)}) \bPsi , \qquad i\ne j,
\end{equation}
where $A_{ij}:\ZZ^N\to\RR$ are some functions constrained by the compatibility
of the system \eqref{eq:4p-affine}. 
It is also convenient \cite{BoKo} to replace the second order linear system 
\eqref{eq:4p-affine} by a first order system as follows.
The compatibility of
\eqref{eq:4p-affine} allows for definition of the potentials (the Lam\'{e}
coefficients) $H_i:\ZZ^N\to\RR$ such that
\begin{equation}
A_{ij} = \frac{H_{i(j)}}{H_i}, \; \; \; i\ne j \; .
\end{equation}
The new wave functions $\bpsi_i$ given by the decomposition
\begin{equation} \label{eq:H-X}
\D_i \bPsi = H_{i(i)} \bpsi_i\; ,
\end{equation}
satisfy the first order linear system
\begin{equation} \label{eq:lin-X}
\Delta_j\bpsi_i = Q_{ij(j)}\bpsi_j,    \; \; \; i\ne j \; ,
\end{equation}
where the functions $Q_{ij}$, called the rotation coefficients, are calculated 
from the
equation
\begin{equation} \label{eq:lin-H}
\D_iH_j =  Q_{ij} H_{i(i)}, \; \; \; i\ne j \; ,
\end{equation}
which is called adjoint to \eqref{eq:lin-X}.
Both the equations \eqref{eq:lin-H} and \eqref{eq:lin-X}
are compatible provided the fields $Q_{ij}$ satisfy the discrete Darboux
equations \cite{BoKo}
\begin{equation} \label{eq:MQL-Q}
\D_kQ_{ij} = Q_{ik(k)} Q_{kj}\;\; , \qquad  i\neq j\neq k\neq i.
\end{equation}
\begin{Cor} \label{cor:4p-psi-i}
Due to \eqref{eq:lin-X} the function $\bpsi_i$ satisfies itself 
the four point equation
\begin{equation} \label{eq:4p-psi-i}
\bpsi_{i(ij)} = \bpsi_{i(i)} + \frac{Q_{ij(ij)}}{Q_{ij(j)}}\bpsi_{i(j)} -
\frac{Q_{ij(ij)}}{Q_{ij(j)}}\left( 1 - Q_{ij(j)}Q_{ji(i)} \right) \bpsi_i,
\qquad i\ne j,
\end{equation}
while the function $\left( \frac{H_j}{Q_{ij}} \right)_{(ij)}$ satifies the
adjoint of \eqref{eq:4p-psi-i} in the sense of \eqref{6ad}.
\end{Cor}
The discrete Darboux equations imply existence of the potentials $\rho_i$ given as
solutions of the compatible system
\begin{equation} \label{eq:rho-constr}
\frac{\rho_{i(j)}}{\rho_i} = 1 - Q_{ji(i)} Q_{ij(j)}, \qquad  i\ne j \; ,
\end{equation}
and yet another potential $\tau$ such that
\begin{equation} \label{eq:tau}
\rho_i = \frac{\tau_{(i)}}{\tau} .
\end{equation}
In terms of the $\tau$-function and the functions
\begin{equation} \label{eq:tau-ij}
\tau_{ij} = \tau Q_{ij} \; ,
\end{equation}
the meaning of which will be given in section \ref{sec:Laplace-tr},
equations \eqref{eq:rho-constr} and
\eqref{eq:MQL-Q} can be rewritten \cite{DMMMS,DS-sym}
in the bilinear form
\begin{align} \label{eq:Hir-ij}
\tau_{(ij)}  \tau & = \tau_{(i)} \tau_{(j)} - \tau_{ji(i)} \tau_{ij(j)} \; ,
\qquad i\ne j\\
 \label{eq:Hir-ijk}
\tau_{ij(k)}  \tau  & = \tau_{(k)} \tau_{ij} + \tau_{ik(k)} \tau_{kj} \;,
\qquad i\ne j \ne k \ne i .
\end{align}

\subsection{The vectorial fundamental (bilinear Darboux) transformation}
\label{sec:fund}
We start with a simple algebraic fact, whose consequences will be discussed
throughout the remaining part of this Section.
\begin{Th}[\cite{MDS}] \label{th:vect-Darb}
Given the solution $\boldsymbol{\phi}_i:\mathbb{Z}^N\to\UU$, 
of the linear system \eqref{eq:lin-X}, and given the solution 
$\boldsymbol{\phi}^*_i:\mathbb{Z}^N\to\WW^*$, of the adjoint linear system 
\eqref{eq:lin-H}. These allow to
construct the linear operator valued potential 
$\boldsymbol{\Omega}[\boldsymbol{\phi},\boldsymbol{\phi}^*]:
\mathbb{Z}^N\to \mathrm{L}(\WW,\UU)$,
defined by 
\begin{equation} \label{eq:Omega-Y-Y}
\Delta_i \boldsymbol{\Omega}[\boldsymbol{\phi},\boldsymbol{\phi}^*] = 
\boldsymbol{\phi}_i \otimes \boldsymbol{\phi}^*_{i(i)}, 
\qquad i = 1,\dots , N.
\end{equation} 
If $\WW=\UU$ and the potential $\boldsymbol{\Omega}$ is invertible, 
$\boldsymbol{\Omega}[\boldsymbol{\phi},\boldsymbol{\phi}^*]\in\mathrm{GL}(\WW)$,
then
\begin{align}
\tilde{\bphi}_i &=\bOm[\bphi,\bphi^*]^{-1}\bphi_i, \;\; \; i=1,...,N,
\label{eq:hatY}\\
\tilde{\bphi}^*_i&=\bphi^*_i\bOm[\bphi,\bphi^*]^{-1},\;\;\; i=1,...,N,\label{eq:hatY*}
\end{align}
satisfy the linear systems \eqref{eq:lin-X}-\eqref{eq:lin-H} correspondingly,
with the fields
\begin{equation}
\tilde{Q}_{ij} = Q_{ij}-\langle \bphi^*_j|\bOm[\bphi,\bphi^*]^{-1}|\bphi_i \rangle,
\;\;\; i,j=1,\dots,N, \;\;i\neq j, \label{eq:hatQ}.
\end{equation}
In addition,
\begin{equation} \label{eq:hatOm}
\bOm[\tilde{\bphi},\tilde{\bphi}^*] = C - \bOm[\bphi,\bphi^*]^{-1},
\end{equation}
where $C$ is a constant operator.
\end{Th}
\begin{Rem}
Notice that because of \eqref{eq:H-X} we have $\Psi = \bOm[\bpsi ,H]$.
\end{Rem}
\begin{Cor}[\cite{MM}]
The potentials $\rho_i$ 
and the $\tau$-function
transform according
to
\begin{align} 
\label{eq:fund-vect-rho}
\tilde{\rho}_i & = \rho_i(1 + \bphi^*_{i(i)}
\boldsymbol{\Omega}[\bphi,\bphi^*]^{-1}
\bphi_i),\\
\tilde{\tau} & = \tau \det\boldsymbol{\Omega}[\bphi,\bphi^*].
\end{align}
\end{Cor}
Applying the above transformation one can produce new compatible (affine) four
point linear problems from the old ones.

To obtain conventional  transformation formulas 
consider \cite{TQL} the following splitting of the vector space $\WW$ of
Theorem \ref{th:vect-Darb}:
\begin{equation} \WW = \EE\oplus\VV\oplus\FF  \;, \; \; \;
 \WW^* = \EE^*\oplus\VV^*\oplus\FF^* \; \; \; \; ,
\end{equation}
if
\begin{equation}
 \bphi_i = \begin{pmatrix} \bpsi_i \\ \btheta_i \\ 0 \end{pmatrix} , \; \; \;
\bphi^*_i = ( 0 , \btheta^*_i, \bpsi^*_i ) \; ,
\end{equation}
then, the corresponding potential matrix is of the form
\begin{equation}
\bOm[\bphi,\bphi^*] = \begin{pmatrix}
\II_\EE & \bOm[\bpsi,\btheta^*] & \bOm[\bpsi, \bpsi^*] \\
0       & \bOm[\btheta,\btheta^*] & \bOm[\btheta, \bpsi^*] \\
0       &      0          &   \II_\FF
\end{pmatrix}
\end{equation}
and its inverse is
\begin{equation}
\bOm[\bphi,\bphi^*]^{-1} =
\begin{pmatrix}
\II_\EE & -\bOm[\bpsi,\btheta^*]  \bOm[\btheta,\btheta^*]^{-1} &
-\bOm[\bpsi, \bpsi^*] +\bOm[\bpsi,\btheta^*]\bOm[\btheta,\btheta^*]^{-1}\bOm[\btheta, \bpsi^*]  \\
0       & \bOm[\btheta,\btheta^*]^{-1} & -  \bOm[\btheta,\btheta^*]^{-1}\bOm[\btheta, \bpsi^*] \\
0       &      0          &   \II_\FF
\end{pmatrix}.
\end{equation}

Let us consider the case of $K$-dimensional transformation data space, $\VV
=\RR^K$, and $\FF =\RR$,
$\EE = \RR^M$, then the transformed solution $\tilde{\bPsi} = \bOm[\tilde{\bpsi}, 
\tilde{H}]$
of the four point scheme (recall that $\bPsi = \bOm[\bpsi,H]$) up to a
constant vector reads
\begin{equation} \label{eq:fund-x}
\tilde{\bPsi} = \bPsi - \bOm[\bpsi, \btheta^*]\bOm[ \btheta, \btheta^*]^{-1}
\bOm[ \btheta, H],
\end{equation}
where the corresponding transformed solutions $\tilde{\bpsi}_i$  of the linear problem
\eqref{eq:lin-X} and $\tilde{H}_i$ of the adjoint linear problem \eqref{eq:lin-H}
are given by equations
\begin{align}
\tilde{\bpsi}_i & = \bpsi_i -\bOm[\bpsi,\btheta^*]  
\bOm[\btheta,\btheta^*]^{-1}\btheta_i ,\\
\tilde{H}_i & = H_i - \btheta^*_i \bOm[\btheta,\btheta^*]^{-1}\bOm[\btheta, H],
\end{align}
and
\begin{equation} \label{eq:Q-fund}
\tilde{Q}_{ij} = Q_{ij} - \btheta^*_j \bOm[\btheta,\btheta^*]^{-1}\btheta_i.
\end{equation}
The scalar ($K=1$) fundamental transformation in the above form
 was given in \cite{KoSchief2}.

\begin{Rem}
To connect the above formalism to the results of Section~\ref{sec:6p} notice
that given scalar solution $\theta_i$ of the linear system \eqref{eq:lin-X} then
the potential $\Theta = \Omega[\theta, H]$ is the scalar solution of the 
system \eqref{eq:4p-affine} of second order linear equations. Moreover, 
given the solution $\theta^*_i$, of the adjoint linear system 
\eqref{eq:lin-H} then the functions 
\begin{equation} \label{eq:sigma-i-def}
\sigma_i = \frac{\theta^*_i}{H_i}
\end{equation}
satisfy \cite{TQL} the linear system
\begin{equation} \label{eq:sigma}
\Delta_j \sigma_i = (A_{ij}-1) \left( \sigma_{j(j)} - \sigma_{i(j)} \right),
\qquad i\ne j.
\end{equation}
Equations \eqref{eq:sigma} imply that the functions 
\begin{equation} \label{eq:Phi-ij}
\Phi_{ij} = \sigma_{i(ij)}
- \sigma_{j(ij)},
\qquad i\ne j,
\end{equation}
satisfy the corresponding equations
\begin{equation} \label{eq:4p-affine-adj-ij}
\Phi_{ij(-i -j)} = A_{ij}\Phi_{ij(-i)} + 
A_{ji} \Phi_{ij(-j)} +( 1- A_{ij(i)} - A_{ji(j)})\Phi_{ij}  , \qquad i\ne j,
\end{equation}
adjoint of equations \eqref{eq:4p-affine}. 
In the case $M=2$ we obtain therefore the data of the transformation being the
solution $\Theta$ of the linear problem and the solution $\Phi=\Phi_{12}$
of its adjoint, thus we
recover results of Theorem~\ref{th:6p-Dt} with the four point 
specification \eqref{4ps} in the affine gauge; see \cite{Dol-Koe} for 
more detailed description. Notice that in order to describe fundamental
transformation in the second order formalism for $M>2$ one should consider in 
addition the algebraic relations
\begin{equation*}
\Phi_{ij} = - \Phi_{ji}, \qquad \Phi_{ij(-i -j)} + \Phi_{jk(-j -k)} =
\Phi_{ik(-i -k)},
\end{equation*}
which are consequences of definition \eqref{eq:Phi-ij}.
\end{Rem}

It is important to notice that the vectorial fundamental transformation 
can be obtained as a
superposition of $K$ scalar transformations,
which follows from the following observation.
\begin{Prop}[\cite{TQL}]
\label{th:perm-fund}
Assume the following splitting of the data of the vectorial fundamental
transformation
\begin{equation}
\btheta_i = \left( \begin{array}{c} 
\btheta_i^a \\ \btheta_i^b \end{array} \right),\qquad
\btheta_i^* = \left( \begin{array}{cc} 
\btheta_{ai}^{*}  & \btheta_{b i}^{*} \end{array} \right),
\end{equation}
associated with the partition $\RR^K = \RR^{K_a} \oplus \RR^{K_b}$,
which implies the following splitting of the potentials
\begin{equation} \label{eq:split-fund-1}
\boldsymbol{\Omega}[\btheta,H] =  \left( \begin{array}{c} 
\boldsymbol{\Omega}[\btheta^a,H] \\ 
\boldsymbol{\Omega}[\btheta^b,H] \end{array} \right), \qquad
\boldsymbol{\Omega}[\btheta,\btheta^*] = \left( \begin{array}{cc}
\boldsymbol{\Omega}[\btheta^a,\btheta_a^*] &
\boldsymbol{\Omega}[\btheta^a,\btheta_b^*] \\
\boldsymbol{\Omega}[\btheta^b,\btheta_a^*] &
\boldsymbol{\Omega}[\btheta^b,\btheta_b^*]\end{array} \right), 
\end{equation}
\begin{equation} \label{eq:split-fund-2}
\boldsymbol{\Omega}[\boldsymbol{\psi},\btheta^*] = \left( \begin{array}{cc} 
\boldsymbol{\Omega}[\boldsymbol{\psi},\btheta_a^*]  &
\boldsymbol{\Omega}[\boldsymbol{\psi},\btheta_b^*]\end{array} \right).
\end{equation} 
Then the vectorial fundamental transformation is equivalent to the following
superposition of vectorial fundamental transformations:\\
1] Transformation $\bPsi\to\bPsi^{\{a\}}$ with the data 
$\btheta_i^a$, $\btheta_{ai}^*$ and the corresponding
potentials
$\boldsymbol{\Omega}[\btheta^a,H]$, 
$\boldsymbol{\Omega}[\btheta^a,\btheta_a^*]$, 
$\boldsymbol{\Omega}[\boldsymbol{\psi},\btheta_a^*]$
\begin{align}
\label{eq:fund-vect-a}
\bPsi^{\{a\}}  & = \bPsi - 
\boldsymbol{\Omega}[\boldsymbol{\psi},\btheta^*_a]
\boldsymbol{\Omega}[\btheta^a,\btheta^*_a]^{-1}
\boldsymbol{\Omega}[\btheta^a,H],\\
\boldsymbol{\psi}_i^{\{a\}} & = \boldsymbol{\psi}_i -
\boldsymbol{\Omega}[\boldsymbol{\psi},\btheta^*_a]
\boldsymbol{\Omega}[\btheta^a,\btheta^*_a]^{-1}
\btheta^a_i,
\\
H_i^{\{a\}} & = H_i - \btheta^*_{i a}
\boldsymbol{\Omega}[\btheta^a,\btheta^*_a]^{-1}
\boldsymbol{\Omega}[\btheta^a,H].
\end{align}
2] Application on the result the vectorial fundamental transformation with the
transformed data
\begin{align}
{\btheta}_i^{b\{a\}} & = \btheta_i^b -
\boldsymbol{\Omega}[\btheta^b,\btheta^*_a]
\boldsymbol{\Omega}[\btheta^a,\btheta^*_a]^{-1}
\btheta^a_i,
\\
{\btheta}_{i b}^{*\{a\}} & = \btheta_{i b}^* - 
\btheta^*_{i a}
\boldsymbol{\Omega}[\btheta^a,\btheta^*_a]^{-1}
\boldsymbol{\Omega}[\btheta^a, \btheta_{b}^*],
\end{align}
and potentials
\begin{align} 
{\boldsymbol{\Omega}}[\btheta^b,H]^{\{a\}} & =
\boldsymbol{\Omega}[\btheta^b,H] - 
\boldsymbol{\Omega}[\btheta^b,\btheta^*_a]
\boldsymbol{\Omega}[\btheta^a,\btheta^*_a]^{-1}
\boldsymbol{\Omega}[\btheta^a,H]=
\boldsymbol{\Omega}[{\btheta}^{b\{a\}},H^{\{a\}}],
\\
{\boldsymbol{\Omega}}[\btheta^b,\btheta^*_b]^{\{a\}} & =
\boldsymbol{\Omega}[\btheta^b,\btheta^*_b] - 
\boldsymbol{\Omega}[\btheta^b,\btheta^*_a]
\boldsymbol{\Omega}[\btheta^a,\btheta^*_a]^{-1}
\boldsymbol{\Omega}[\btheta^a,\btheta^*_b]=
\boldsymbol{\Omega}[{\btheta}^{b\{a\}},{\btheta}_b^{*\{a\}}],
\\
{\boldsymbol{\Omega}}[\boldsymbol{\psi},\btheta^*_b]^{\{a\}}  & =
\boldsymbol{\Omega}[\boldsymbol{\psi},\btheta^*_b] - 
\boldsymbol{\Omega}[\boldsymbol{\psi},\btheta^*_a]
\boldsymbol{\Omega}[\btheta^a,\btheta^*_a]^{-1}
\boldsymbol{\Omega}[\btheta^a,\btheta^*_b]=
\boldsymbol{\Omega}[{\boldsymbol{\psi}}^{\{a\}},{\btheta}_b^{*\{a\}}],
\label{eq:fund-vect-potentials-slit}
\end{align}
i.e.,
\begin{equation} \label{eq:fund-vect-a-b}
\tilde\bPsi = \bPsi^{\{a,b\}}  = \bPsi^{\{a\}} - 
{\boldsymbol{\Omega}}[\boldsymbol{\psi},\btheta^*_b]^{\{a\}}
[{\boldsymbol{\Omega}}[\btheta^b,\btheta^*_b]^{\{a\}}]^{-1}
{\boldsymbol{\Omega}}[\btheta^b,H]^{\{a\}}.
\end{equation}
\end{Prop}
\begin{Rem}
The above formulas, 
apart from existence of the $\tau$-function, remain valid (eventually one needs
the proper ordering of some factors) 
if, instead of the real field $\RR$, we consider \cite{Dol-GAQL} 
arbitrary division ring. Notice, that because the structure of the
transformation formulas \eqref{eq:fund-x} is a consequence of equation
\eqref{eq:hatOm} then the formulas may be expressed in terms of
quasi-determinants \cite{GGRW-quasideterminants} (recall, that roughly speaking,
a quasi-determinant is the inverse of 
an element of the inverse of a matrix with entries in a
division ring).
\end{Rem}
\subsection{Reductions of the fundamental transformation}
Let us list basic reductions of the (scalar)
fundamental transformation. We follow the
nomenclature of \cite{TQL} which has origins in geometric terminology of
transformations of conjugate nets 
\cite{Jonas,DarbouxIV,Eisenhart-TS,Tzitzeica,Bianchi,Lane,Finikov}. 
We provide also the 
terminology of modern theory of integrable systems
\cite{ms,OevelSchief}, where the fundamental transformation is called 
the binary
Darboux transformation. 
All the transformations presented 
in this section can be derived \cite{TQL} from the
fundamental transformation through limiting procedures.

\subsubsection{The L\'{e}vy (elementary Darboux) transformation}
Given a scalar solution $\theta_i$  of the linear problem
\eqref{eq:lin-X}
the L\'evy transform of $\bPsi=\bOm[\bpsi,H]$ is then given by
\begin{equation}
\cL_i(\bPsi) = \bPsi - \frac{\bOm[\theta,H]}{\theta_i} \bpsi_i.
\end{equation}
The transformed Lam\'e coefficients and new wave functions are of
the form
\begin{align}
\cL_i(H_i) & = \frac{1}{\theta_i}\bOm[\theta,H] \; ,\\
\cL_i(H_j) & = H_j - \frac{Q_{ij}}{\theta_i}\bOm[\theta,H] \; , \\
\cL_i(\bpsi_i) & = -\bpsi_{i(i)} + \frac{\theta_{i(i)}}{\theta_i}\bpsi_i \; ,\\
\cL_i(\bpsi_j) & = \bpsi_j - \frac{\theta_j}{\theta_i}\bpsi_i \; .
\end{align}
Within the nonlocal $\bar\partial$-dressing method the elementary Darboux
transformation was introduced in \cite{BoKo}. 

\subsubsection{The adjoint L\'{e}vy (adjoint
elementary Darboux) transformation}
Given a scalar solution $\theta^*_i$  of the adjoint linear problem
\eqref{eq:lin-H}
the adjoint L\'evy transform $\cL^*_i(\bPsi)$ of 
$\bPsi$ is given by
\begin{equation}
\cL^*_i(\bPsi) = \bPsi - \frac{H_i}{\theta^*_i}\bOm[\bpsi,\theta^*].
\end{equation}
The new Lam\'e coefficients and the wave functions are of
the form
\begin{align}
\cL^*_i(H_i) & = H_{i(-i)}- \frac{\theta^*_{i(-i)}}{\theta_i^*}H_i\; ,\\
\cL^*_i(H_j) & = H_j - \frac{\theta_j^*}{\theta_i^*}H_i \; , \\
\cL^*_i(\bpsi_i) & =  \frac{1}{\theta_i^*}\bOm[\bpsi,\theta^*] \; ,\\
\cL^*_i(\bpsi_j) & = \bpsi_j - \frac{Q_{ji}}{\theta_i^*}\bOm[\bpsi,\theta^*] \; .
\end{align}

As it was shown in \cite{TQL}, the scalar fundamental transformation can be
obtained as superposition of the L\'{e}vy transformation and its adjoint. The
closed formulae for iterations of the L\'{e}vy transformations in terms of
Casorati determinants, and analogous result for the adjoint L\'{e}vy
transformation, was given in \cite{LiuManas}.

\begin{Rem}
The description of the L\'{e}vy transformation and its adjoint in the
homogeneous formalism in the case of $M=2$ is given in \cite{Dol-Koe}.
\end{Rem}

\begin{Rem} From analytic \cite{BoKo} and geometric \cite{TQL} point of view 
one can distinguish also the so called \emph{Combescure transformation}, 
whose algebraic description is however
very simple (the wave functions $\bpsi_i$ are invariant). The Combescure
transformation supplemented by the \emph{projective (or radial} 
transformation,
whose algebraic description is also trivial \cite{TQL}), generate the 
fundamental
transformation. See also section \ref{dodane} and \ref{Gc} for 
generalization of the Combescure transformation.
\end{Rem}

\subsubsection{The Laplace (Schlesinger) transformation}
\label{sec:Laplace-tr}
The following transformations does not involve any functional parameters, and
can be considered as further degeneration of the L\'{e}vy (or its adjoint)
reduction.
The Laplace transformation of $\bPsi$ is given by
\begin{equation} \label{def:Lij}
\cL_{ij}(\bPsi) = \bPsi -\frac{H_j}{Q_{ij}}\bpsi_i , \qquad i\ne j.
\end{equation}
The Lam\'e coefficients of the transformed
linear problems read
\begin{align}
\cL_{ij}(H_i) &=  \frac{H_j}{Q_{ij}}
\; , \label{eq:L-Hi} \\
\cL_{ij}(H_j) &= 
\left( Q_{ij}\D_j \left(\frac{H_j}{Q_{ij}}  \right) \right)_{(-j)}
\; , \label{eq:L-Hj} \\
\cL_{ij}(H_k) &= H_k - \frac{Q_{ik}}{Q_{ij}} H_j
\; , \; \; k\ne i,j, \label{eq:L-Hk}
\end{align}
and the new wave functions read
\begin{align}
\cL_{ij}(\bpsi_i) & = - \D_i\bpsi_i + \frac{\D_iQ_{ij}}{Q_{ij}} \bpsi_i
\; ,\label{eq:L-Xi} \\
\cL_{ij}(\bpsi_j) & = -\frac{1}{Q_{ij}}\bpsi_i \; ,
\label{eq:L-Xj} \\
\cL_{ij}(\bpsi_k) & = \bpsi_k - \frac{Q_{kj}}{Q_{ij}} \bpsi_i
\; ,\; \; k\ne i,j \label{eq:L-Xk} \; .
\end{align}
The Laplace transformations satisfy generically the following identities
\begin{align}
\label{eq:Lid-ij}
\cL_{ij} \circ \cL_{ji} & = \text{id} \; ,\\
\cL_{jk} \circ \cL_{ij} &= \cL_{ik}, \\
\cL_{ki} \circ \cL_{ij} &= \cL_{kj}. \label{eq:Lid-ijk}
\end{align}
The Laplace transformation for the four point affine
scheme was introduced in
\cite{DCN} following the geometric ideas of \cite{Sauer} and  independently in 
\cite{NovVes} using the factorization approach. 
The generalization for systems of four point schemes
(quadrilateral lattices) was given in \cite{TQL}.
\begin{Rem}
As it was shown in \cite{Dol-Hir} the functions $\tau_{ij}$ defined by equation
\eqref{eq:tau-ij} are $\tau$-functions of the transformed four point schemes $\cL_{ij}(\bPsi)$
\begin{equation}
\tau_{ij} = \cL_{ij}(\tau),
\end{equation} 
which, due to \eqref{eq:Lid-ij}, leads \cite{DCN} 
to the discrete Toda system \cite{Hirota}.
\end{Rem}

%%%%%%%%%%%%%%%%%%%%%%%%%%%%%%%%%%%%%%%%%%%%%%%%%%%%%%%%%%%%%%%
\section{Distinguished reductions of the four point scheme} 
%%%%%%%%%%%%%%%%%%%%%%%%%%%%%%%%%%%%%%%%%%%%%%%%%%%%%%%%%%%%%
\label{sec:4p-red}
In this section we study (systems of) four point linear equations subject to
additional constraints, and we provide corresponding reductions of the
fundamental transformation. The basic algebraic
idea behind such reduced transformations
lies in a relationships between solutions of the linear problem and its adjoint,
which should be preserved by the fundamental transformation (see, for example, 
application of this technique in \cite{NimmoWillox}
to reductions of the binary Darboux transformation for the Toda
system). Some results presented here have been partially covered in Section 
\ref{sec:4p-6p} but in a different setting.

\subsection{The Moutard (discrete BKP) reduction}
\label{bkp}
Consider the system of discrete Moutard equations (the
discrete BKP linear problem \cite{DJM,NiSchief})
\begin{equation} \label{eq:BKP-lin}
\bPsi_{(ij)} - \bPsi = F_{ij} (\bPsi_{(i)} - \bPsi_{(j)}) , \quad 1\leq i< j\leq N,
\end{equation}
for suitable functions $F_{ij}:\ZZ^N\to\RR$. Compatibility of the system  
implies existence of the potential $\tau^B:\ZZ^N\to\RR$, 
in terms of which the functions $F_{ij}$ can be written as
\begin{equation} \label{eq:tau-B}
F_{ij} = \frac{\tau^B_{(i)}\tau^B_{(j)}}{\tau^B \, \tau^B_{(ij)}}, 
\qquad i\ne j ,
\end{equation}
which satisfies
system of Miwa's discrete 
BKP equations \cite{Miwa}
\begin{equation} \label{eq:BKP-nlin}
\tau^B \tau^B_{(ijk)} = \tau^B_{(ij)}\tau^B_{(k)} - \tau^B_{(ik)}\tau^B_{(j)} + 
\tau^B_{(jk)}\tau^B_{(i)}, \quad 1\leq i< j < k \leq N.
\end{equation}
The discrete Moutard system can be given \cite{BQL} the first order formulation
\eqref{eq:lin-X}-\eqref{eq:lin-H} upon introducing the Lam\'{e} coefficients
\begin{equation} \label{eq:H-tau-B}
H_i = (-1)^{\sum_{k<i}m_k}\frac{\tau^B_{(-i)}}{\tau^B},
\end{equation}
and the rotation coefficients (below we assume $i<j$)
\begin{align} \label{eq:Q-BQL-ij}
Q_{ij} & = - (-1)^{\sum_{i\leq k<j}m_k}
\left( \frac{\tau^B_{(i-j)}}{\tau^B_{(-j)}} + \frac{\tau^B_{(i)}}{\tau^B}\right) 
\frac{\tau^B_{(-j)}}{\tau^B},\\
\label{eq:Q-BQL-ji}
Q_{ji} & = - (-1)^{\sum_{i\leq k<j}m_k}
\left( \frac{\tau^B_{(-i j)}}{\tau^B_{(-i)}} - \frac{\tau^B_{(j)}}{\tau^B}\right)
\frac{\tau^B_{(-i)}}{\tau^B},
\end{align}
which in view of \eqref{eq:rho-constr},
gives the familiar relation between the $\tau$-functions of the KP
and BKP hierarchies
\begin{equation}
\tau = \left(\tau^B\right)^2.
\end{equation}
The corresponding reduction of the fundamental transformation was given in
\cite{BQL}, where also a link with earlier work \cite{NiSchief}
 on the discrete Moutard transformation has been established.
\begin{Prop}[\cite{BQL}] \label{prop:vect-fund-red-BQL}
Given solution $\boldsymbol{\theta}_i:\ZZ^N\to\RR^K$ of the linear problem
\eqref{eq:lin-X} corresponding to the Moutard linear system \eqref{eq:BKP-lin}
and its first order form \eqref{eq:H-tau-B}-\eqref{eq:Q-BQL-ji}.
Denote by $\boldsymbol{\Theta} = \boldsymbol{\Omega}[\boldsymbol{\theta},H]$ the
corresponding potential, which is also new
vectorial solution of the linear problem \eqref{eq:BKP-lin}.\\ 
1) Then
\begin{equation} \label{eq:BQL-Y*}
\boldsymbol{\theta}^*_i = (-1)^{\sum_{k<i}m_k}\frac{\tau^B_{(-i)}}{\tau^B} 
(\boldsymbol{\Theta}_{(-i)}^T + \boldsymbol{\Theta}^T)
\end{equation}
provides a vectorial
solution of the adjoint linear problem, and the corresponding 
potential
$\boldsymbol{\Omega}[\boldsymbol{\theta},\boldsymbol{\theta}^*]$ allows 
for the following
constraint
\begin{equation} \label{eq:vect-Moutard-constr}
\boldsymbol{\Omega}[\boldsymbol{\theta},\boldsymbol{\theta}^*] + 
\boldsymbol{\Omega}[\boldsymbol{\theta},\boldsymbol{\theta}^*]^T =
2 \boldsymbol{\Theta}\otimes \boldsymbol{\Theta}^T.
\end{equation}
2) The fundamental vectorial transform $\tilde\bPsi$ of $\bPsi$, given by
\eqref{eq:fund-x} with the potentials $\boldsymbol{\Omega}$ restricted as
above satisfies Moutard linear system \eqref{eq:BKP-lin} and 
can be considered as the superposition of $K$ scalar
reduced fundamental transforms. 
\end{Prop}
\begin{Rem}
In the scalar case and for $N=2$ inserting in equations \eqref{eq:sigma-i-def}
and \eqref{eq:sigma}
the functions $H_i$ and $\theta^*i$ given by \eqref{eq:H-tau-B} and
\eqref{eq:BQL-Y*} correspondingly, we obtain the formula \eqref{PT} in the gauge
$F=1$ (modulo the corresponding change of sign).
\end{Rem}
Notice that given $\boldsymbol{\Theta}$ then, because of the constraint
\eqref{eq:vect-Moutard-constr}, to construct 
$\boldsymbol{\Omega}(\boldsymbol{\theta},\boldsymbol{\theta}^*)$ 
we need only its antisymmetric part 
$S(\boldsymbol{\Theta} | \boldsymbol{\Theta})$, which satisfies the system
\begin{equation}
\Delta_i S(\boldsymbol{\Theta} | \boldsymbol{\Theta}) = 
\boldsymbol{\Theta}_{(i)}\otimes \boldsymbol{\Theta}^{\mathrm{t}} -
\boldsymbol{\Theta}\otimes \boldsymbol{\Theta}^{\mathrm{t}}_{(i)}.
\end{equation} 
This observation is the key element of the connection of the above reduction of
the fundamental transformation with earlier results \cite{NiSchief} on the
vectorial Moutard transformation for the system \eqref{eq:BKP-lin}, where the
formulas using Pfaffians were obtained (recall that determinant of a
skew-symmetric matrix is a square of Pfaffian). In particular, the 
transformation
rule for the $\tau^B$-function can be recovered 
\begin{equation}
\tilde\tau^B = \begin{cases}
\tau^B \: \mathrm{Pf} \: S(\boldsymbol{\Theta} | \boldsymbol{\Theta}) , & 
K \; \text{even} \\
\tau^B \: \mathrm{Pf} \left( \begin{array}{cc}
0 & - \boldsymbol{\Theta}^T , \\
\boldsymbol{\Theta} &  S(\boldsymbol{\Theta} | \boldsymbol{\Theta})
\end{array} \right), \qquad & K \; \text{odd}.
\end{cases}
\end{equation}

\subsection{The symmetric (discrete CKP) reduction}
\label{ckp}
Consider the linear problem subject to the constraint \cite{DS-sym} which arose
from studies on the Egorov lattices \cite{Schief-talk}
\begin{equation} \label{eq:symm}
\rho_i Q_{ji(i)}=\rho_j Q_{ij(j)} \; , \quad i\ne j \; .
\end{equation}
Then the discrete Darboux equations \eqref{eq:MQL-Q} can be 
rewritten~\cite{Schief-JNMP} in the following quartic form
\begin{equation} \begin{split}
\left( \tau   \tau_{(ijk)} - \tau_{(i)}  \tau_{(jk)} -
\tau_{(j)}  \tau_{(ik)} -  \tau_{(k)}  \tau_{(ij)} \right)^2 
+ 4(\tau_{(i)}\tau_{(j)}\tau_{(k)}\tau_{(ijk)} + &
\tau  \tau_{(ij)}\tau_{(jk)} \tau_{(ik)} )= \\
4( \tau_{(i)}  \tau_{(jk)} \tau_{(j)}  \tau_{(ik)} + \tau_{(i)}  \tau_{(jk)} 
\tau_{(k)}  \tau_{(ij)}  + 
\tau_{(j)}  \tau_{(ik)}\tau_{(k)}  \tau_{(ij)}), &
\end{split} \end{equation}
which can be identified with equation derived in \cite{Kashaev} in connection 
with the star-triangle relation in the
Ising model. According to~\cite{Schief-JNMP}, the above equation can be obtained
from the CKP hierarchy via successive application of the corresponding reduction
of the binary Darboux transformations.

Construction~\cite{MM} of the reduction of the fundamental transformations 
which preserves
the constraint \eqref{eq:symm}
makes use the following observation. 
\begin{Lem}[\cite{DS-sym}]
The following conditions are equivalent:\\
1) The functions $Q_{ij}$, $\rho_i$ satisfy constraint \eqref{eq:symm};\\
2) Given a nontrivial solution $\bphi_i^*$ of the adjoint linear problem 
\eqref{eq:lin-H} then 
\begin{equation} \label{eq:C-ft-YY}
\bphi_i = \rho_i (\bphi^*_{i(i)})^T
\end{equation}
provides a solution of the linear problem \eqref{eq:lin-X}; \\
3) The corresponding potential
$\boldsymbol{\Omega}[\boldsymbol{\phi},\boldsymbol{\phi}^*] $ allows for the 
constraint
\begin{equation}\label{eq:C-ft-O}
\boldsymbol{\Omega}[\boldsymbol{\phi},\boldsymbol{\phi}^*]^T=
\boldsymbol{\Omega}[\boldsymbol{\phi},\boldsymbol{\phi}^*].
 \end{equation}
\end{Lem}
\begin{Th}[\cite{MM}] 
When the data $\btheta_i$, $\btheta^*_i$, $\bOm[\btheta,\btheta^*]$ of the fundamental
transformation satisfy conditions
\eqref{eq:C-ft-YY}-\eqref{eq:C-ft-O} then the new functions $\tilde{Q}_{ij}$, 
$\tilde{\rho}_i$ found by equations \eqref{eq:Q-fund}, \eqref{eq:fund-vect-rho}
are constrained by \eqref{eq:symm}, i.e.
\begin{equation*}
\tilde{\rho}_i \tilde{Q}_{ji(i)}=\tilde{\rho}_j \tilde{Q}_{ij(j)} \; , \quad i\ne j \; .
\end{equation*}
\end{Th}
The corresponding permutability principle has been proved in \cite{CQL}.
\begin{Rem}
To connect the symmetric reduction with the Goursat equation and the
corresponding transformation (see Section~\ref{Gr}) notice that because of
Corollary~\ref{cor:4p-psi-i} and equations \eqref{eq:rho-constr}-\eqref{eq:tau}
the function $\bpsi_2$ satisfies equation \eqref{dG}
with $q=Q_{21(1)}$. Then, due to the same Corollary and condition
\eqref{eq:C-ft-YY}, given scalar solution $\theta_2$ of the four point equation of
$\bpsi_2$ then the function 
\begin{equation*}
\left(\frac{\theta_1}{\rho_2
Q_{21(1)}}\right)_{(2)} = \left(\frac{\tau \Delta_1 \theta_2}{\tau_{(2)}
Q_{21(1)}^2}\right)_{(2)},  
\end{equation*} 
satisfies its adjoint; in the lat equation we have used the linear problem
\eqref{eq:lin-X} and equation \eqref{eq:tau}.
\end{Rem}

\subsection{Quadratic reduction}
\label{Qr}
Consider the system of four point equations \eqref{eq:4p-affine}
which solution $\bPsi$ is subject to the following quadratic constraint
\begin{equation} \label{eq:quadric}
\bPsi^T  Q  \bPsi +  \ba^T \bPsi + c = 0  \; \; \; ;
\end{equation}
here $Q$ is a non-degenerate symmetric matrix, $\ba$
is a constant vector, $c$ is a scalar.
\begin{Rem}
Notice that unlike in two previous reductions we fix (by giving the
quadratic equation) the dimension of $\bPsi$.
\end{Rem}
Double discrete differentiation of equation \eqref{eq:quadric} in $i\ne j$
directions gives, after some algebra, the condition
\begin{equation} \label{eq:XQX}
\bpsi_{i(j)}^T  Q  \bpsi_j +  \bpsi_{j(i)}^T  Q  \bpsi_i  = 0,
\end{equation}
analogous to that obtained in \cite{DMS} in order to characterize circular
lattices~\cite{Bobenko-DO,CDS}. It implies \cite{DS-sym}, in particular, that
$\bpsi_i^T  Q  \bpsi_i$ satisfy  the same equation \eqref{eq:rho-constr}
as the potentials $\rho_i$.

As in two above reductions, the quadratic condition allows for a relation
between solutions of the linear system \eqref{eq:lin-X} and its adjoint
\eqref{eq:lin-H}. The following proposition can be easily derived from analogous
results of \cite{q-red}, where as the basic ingredient of the transformation was used
the
potential $\bOm[\bpsi, \btheta^*]$, but we present here its direct proof in
the spirit of corresponding results found for the circular
lattice~\cite{KoSchief2,LiuManas-SR}.
\begin{Prop}
Given a nontrivial solution $\btheta_i^*$ of the adjoint linear problem 
\eqref{eq:lin-H} corresponding to the system of four point equations 
\eqref{eq:4p-affine} which solution $\bPsi$ is subject to the constraint
\eqref{eq:quadric} then
\begin{equation} \label{eq:Q-ft-YY}
\btheta_i = (\bOm[\bpsi, \btheta^*]_{(i)} + \bOm[\bpsi, \btheta^*])^T Q \bpsi_i
\end{equation}
provides a solution of the corresponding linear problem \eqref{eq:lin-X}. 
\end{Prop}
\begin{proof}
After some algebra using the equations satisfied by $\bY^*_i$ and $\bpsi_i$ one
gets
\begin{equation}
\D_j \btheta_i - Q_{ij(j)}\btheta_j = \btheta^{*T}_{j(ij)}\left(
\bpsi_{i(j)}^T  Q  \bpsi_j +  \bpsi_{j(i)}^T  Q  \bpsi_i \right),
\end{equation}
which vanishes due to \eqref{eq:XQX}.
\end{proof}
The following result gives the discrete Ribaucour reduction of the fundamental
transformation.
\begin{Prop}[\cite{q-red}]
Given solution $\boldsymbol{\theta}^*_i:\ZZ^N\to\RR^K$ of the adjoint
linear problem \eqref{eq:lin-H} corresponding to the quadratic constraint
\eqref{eq:quadric}. \\
1) Then the potentials $\bOm[\bpsi, \btheta^*]$, $\bOm[\btheta, H]$ and
$\bOm[\btheta, \btheta^*]$, where $\btheta_i$ is the solution of the linear problem
\eqref{eq:lin-X} constructed from $\btheta^*_i$ by means of formula
\eqref{eq:Q-ft-YY}, allow for the constraints
\begin{align}
\label{eq:bphi-qv}
\bOm[\btheta,H]^T & = 2 \bpsi^T Q  \bOm[\bpsi, \btheta^*]  +
 \ba^T \bOm[\bpsi, \btheta^*]  \; \; , \\
\label{eq:bPhiC-qv}
\bOm[\btheta, \btheta^*] + \bOm[\btheta, \btheta^*]^T & = 
 2 \bOm[\bpsi, \btheta^*]^T  Q \bOm[\bpsi, \btheta^*]  \; .
\end{align}
2) The Ribaucour reduction of the
fundamental vectorial transform $\tilde\bPsi$ of $\bPsi$, given by
\eqref{eq:fund-x} with the potentials $\boldsymbol{\Omega}$ restricted as
above satisfies the quadratic constraint \eqref{eq:quadric} and 
can be considered as the superposition of $K$ scalar
Ribaucour transformations. 
\end{Prop}
As it was explained in \cite{q-red}, the Ribaucour transformations
\cite{KoSchief2} of the circular lattice  \cite{Bobenko-DO,CDS} can be derived
from the above approach after the stereographic projection from the M\"{o}bius
sphere. The  superposition principle for
the Ribaucour transformation of circular lattices was derived also in
\cite{LiuManas-SR}.

\section{The three point scheme}
%%%%%%%%%%%%%%%%%%%%%%%%%%%%%%%%%%%%%%%%%%%%%%
\label{3p-sch}
In this final Section we present the vectorial
Darboux transformations for the three
point scheme \eqref{eq:lin-dKP}. The corresponding nonlinear difference system \eqref{eq:H-M},  known as the 
Hirota--Miwa equation, is perhaps the most important and widely studied 
integrable discrete system. It was discovered by Hirota \cite{Hirota}, who called it 
the discrete analogue of the two dimensional Toda lattice (see also \cite{LPS}),
as a culmination of his studies on the bilinear form of 
nonlinear integrable equations. General feature of Hirota's
equation was uncovered by Miwa \cite{Miwa} who found a remarkable transformation
which connects the equation to the KP hierarchy \cite{DKJM}. The Hirota-Miwa
equation, called also the discrete KP equation, can be encountered in various
branches of theoretical physics \cite{Saito,KNS,Zabrodin} and mathematics
\cite{Shiota,KWZ,Octahedron}.

Consider the linear system \cite{DJM-II}
\begin{equation} \label{eq:lin-dKP}
\bPsi_{(i)} - \bPsi_{(j)} = U_{ij} \bPsi ,  \qquad i \ne j \leq N,
\end{equation}
whose compatibility leads to the following parametrization of the field $U_{ij}$
in terms of the potentials $r_i$
\begin{equation}
r_{i(j)} = r_i U_{ij}, 
\end{equation}
and then to existence of the $\tau$ function
\begin{equation}
r_i = (-1)^{\sum_{k<i}m_k}\frac{\tau^H_{(i)}}{\tau^H}
\end{equation} 
and, finally to the the discrete KP system \cite{Hirota,Miwa}
\begin{equation} \label{eq:H-M}
\tau^H_{(i)}\tau^H_{(jk)} - \tau^H_{(j)}\tau^H_{(ik)} + \tau^H_{(k)}\tau^H_{(ij)},
\qquad i< j <k.
\end{equation}
The same nonlinear systems arises from compatibility of
\begin{equation} \label{eq:lin-dKP*}
\bPsi^*_{(j)} - \bPsi^*_{(i)} = 
U_{ij} \bPsi^*_{(ij)} , 
\qquad i \ne j ,
\end{equation}
called the adjoint of \eqref{eq:lin-dKP}.

We present the Darboux transformation for the three point scheme 
in the way similar to that of Section \ref{sec:fund} 
following the approach of \cite{Nimmo-KP}, see however early works on the
subject
\cite{Saito-Saitoh,Saitoh-Saito}.
\begin{Th} \label{th:vect-Darb-KP}
Given the solution $\boldsymbol{\Phi}:\mathbb{Z}^N\to\UU$, 
of the linear system \eqref{eq:lin-dKP}, and given the solution 
$\boldsymbol{\Phi}^*:\mathbb{Z}^N\to\WW^*$, of the adjoint linear system 
\eqref{eq:lin-dKP*}. These allow to
construct the linear operator valued potential 
$\boldsymbol{\Omega}[\boldsymbol{\Phi},\boldsymbol{\Phi}^*]:
\mathbb{Z}^N\to \mathrm{L}(\WW,\UU)$,
defined by 
\begin{equation} \label{eq:omega-p-p}
\Delta_i \boldsymbol{\Omega}[\boldsymbol{\Phi},\boldsymbol{\Phi}^*] = 
\boldsymbol{\Phi} \otimes \boldsymbol{\Phi}^*_{(i)}, 
\qquad i = 1,\dots , N.
\end{equation} 
If $\WW=\UU$ and the potential $\boldsymbol{\Omega}$ is invertible, 
$\boldsymbol{\Omega}[\boldsymbol{\Phi},\boldsymbol{\Phi}^*]\in\mathrm{GL}(\WW)$,
then
\begin{align}
\tilde{\boldsymbol{\Phi}} &=
\bOm[{\boldsymbol{\Phi}},{\boldsymbol{\Phi}}^*]^{-1}\boldsymbol{\Phi}, 
\label{eq:hatp}\\
\tilde{\boldsymbol{\Phi}}^*&=\boldsymbol{\Phi}^*
\bOm[{\boldsymbol{\Phi}},{\boldsymbol{\Phi}}^*]^{-1},\label{eq:hatp*}
\end{align}
satisfy the linear systems \eqref{eq:lin-dKP} and \eqref{eq:lin-dKP*}, 
correspondingly,
with the fields
\begin{equation}
\tilde{U}_{ij} = U_{ij} - 
(\boldsymbol{\Phi}^* \bOm[\boldsymbol{\Phi},\boldsymbol{\Phi^*}]^{-1} 
\boldsymbol{\Phi})_{(i)}
+ (\boldsymbol{\Phi}^* \bOm[\boldsymbol{\Phi},\boldsymbol{\Phi}^*]^{-1}
\boldsymbol{\Phi})_{(j)}.
\end{equation}
In addition,
\begin{equation} \label{eq:hat-om}
\bOm[\tilde{\boldsymbol{\Phi}},\tilde{\boldsymbol{\Phi}}^*] = C - 
\bOm[\boldsymbol{\Phi},\boldsymbol{\Phi}^*]^{-1},
\end{equation}
where $C$ is a constant operator.
\end{Th}
The transformation rule for the potentials $r_i$ reads
\begin{equation}
\tilde{r}_i = r_i (1 - \boldsymbol{\Phi}^*_{(i)} 
\bOm[\boldsymbol{\Phi},\boldsymbol{\Phi}^*]^{-1} \boldsymbol{\Phi} ),
\end{equation}
while using the technique of the bordered determinants \cite{Hirota-book} one can 
show that \cite{Nimmo-KP}
\begin{equation}
\tilde{\tau}^H = \tau^H \det \boldsymbol{\Omega}
[\boldsymbol{\Phi},\boldsymbol{\Phi}^*]  \label{eq:hat-tau-dKP}.
\end{equation}

More standard transformation formulas arise, when one splits, like in 
Section~\ref{sec:fund}, the vector space $\WW$ of
Theorem \ref{th:vect-Darb-KP} as follows
\begin{equation} \WW = \EE\oplus\VV\oplus\FF  \;, \; \; \;
 \WW^* = \EE^*\oplus\VV^*\oplus\FF^* \; \; \; \; ;
\end{equation}
if
\begin{equation}
 \boldsymbol{\Phi} = \begin{pmatrix} 
 \boldsymbol{\Psi} \\ \bTheta \\ 0 \end{pmatrix} , \; \; \;
\boldsymbol{\Phi}^* = ( 0 , \bTheta^*, \boldsymbol{\Psi}^* ) \; .
\end{equation}
Then the corresponding potential matrix and its inverse have the structure like
those in  Section~\ref{sec:fund}, which gives \cite{Nimmo-KP}
\begin{align}
\tilde{\boldsymbol{\Psi}} & = \boldsymbol{\Psi} -
\bOm[\boldsymbol{\Psi},\bTheta^*]  \bom[\bTheta,\bTheta^*]^{-1}\bTheta ,\\
\tilde{\boldsymbol{\Psi}}^* & = \boldsymbol{\Psi}^* - 
\bTheta^* \bOm[\bTheta,\bTheta^*]^{-1}\bOm[\bTheta, \boldsymbol{\Psi}^*],\\
\tilde{U}_{ij}&  = U_{ij} - (\bTheta^* \bOm[\bTheta,\bTheta^*]^{-1} \bTheta)_{(i)}
+ (\bTheta^* \bOm[\bTheta,\bTheta^*]^{-1} \bTheta)_{(j)}.
\end{align}
Notice that one can consider \cite{FWN,FWN-Capel,Nimmo-NCKP} the three point linear problem
in associative algebras. Then the structure of the transformation formulas 
\eqref{eq:hat-om} implies the quasi-determinant interpretation \cite{GNO}
of the above equations.
\begin{Rem}
The permutability property for the Darboux transformations of the three point
scheme \cite{Nimmo-NCKP} can be formulated exactly (cancel subscripts $i$)
like Theorem~\ref{th:perm-fund}.
\end{Rem}

Finally we remark that the 
binary Darboux transformation for the three point linear problem can be
decomposed \cite{Nimmo-KP} into
superposition of the elementary Darboux transformation and its adjoint, which
can be described as follows.
Given a scalar solution $\Theta$  of the linear problem
\eqref{eq:lin-dKP} then
the elementary Darboux transformation
\begin{align}
\cD(\boldsymbol{\Psi}) &= \boldsymbol{\Psi}_{(i)} - \frac{\Theta_{(i)}}{\Theta}
\boldsymbol{\Psi}, \\
\cD(\boldsymbol{\Psi}^*) &= \frac{1}{\Theta}\bOm[\Theta,\boldsymbol{\Psi}^*]\; ,\\
\cD(\tau) &= \Theta \tau
\end{align}
leaves equations \eqref{eq:lin-dKP} and \eqref{eq:lin-dKP*} invariant.

Analogously, given a scalar solution $\Theta^*$  of the linear problem
\eqref{eq:lin-dKP*} then
the adjoint elementary Darboux 
\begin{align}
\cD^*(\boldsymbol{\Psi}) &=\frac{1}{\Theta^*}\bOm[\Theta^*,\boldsymbol{\Psi}],
 \\
\cD^*(\boldsymbol{\Psi}^*) &=  - \boldsymbol{\Psi}^*_{(-i)} + 
\frac{\Theta^*_{(-i)}}{\Theta^*}
\boldsymbol{\Psi}^*, \\
\cD^*(\tau) &= \Theta^* \tau
\end{align}
leaves equations \eqref{eq:lin-dKP} and \eqref{eq:lin-dKP*} invariant.

The above transformations allow for vectorial forms, which can be conveniently
written \cite{Nimmo-KP,OHTI} in terms of
Casorati determinants.

\section{Summary and open problems}
In the paper we aimed to present results on Darboux transformations of 
linear operators on two dimensional regular lattices. To put some 
order into the review we started from the six
point scheme (and its Darboux transformation) as the master linear problem. The
path between its  various specifications and reductions has been visualized in
Figure~\ref{5p}. We considered also in more detail the corresponding 
theory of the Darboux transformations of the systems of the four point schemes
and their reductions. Finally, we briefly discussed the multidimensional
aspects of the three point schemes. It is worth to mention
that for systems of the three or four point schemes the Darboux transformations
can be interpreted as a way to generate new dimensions. This is very much
connected to the permutability of the transformations, which is a core of
integrability of the corresponding nonlinear systems.

Separate issue touched here is such extention of four point schemes  
that can be regarded as an analogue of discretization of a 
diferential equation in arbitrary parametrization.
More precisely we discussed here such an extension of general (AKP) case 
(section \ref{Gc})
and Moutard--selfadjoint (BKP)  case (sections \ref{Mr} and \ref{sec:s-a}). 
Goursat and Ribaucour reductions have not been investigated from this 
point of view.
The multidimensional schemes that mimic equations governing conjugate 
nets (and their reductions) with arbitrary change of independend 
variables have been not exploited either. 

We also mentioned another approach to the problem of construction of the unified
theory of the Darboux transformations. It consists in isolating basic "bricks"
in order to use their combinations to construct more involved linear operators
together with their Darboux transformations. Such an idea has been applied to
derive, starting from the Moutard reduction of the four point scheme, 
the self-adjoint operators on the square (the
five point scheme), triangular (the seven point scheme) 
and the honeycomb grids. Recently it was shown in \cite{DODO} that
the theory of systems of four point 
linear equations and their Laplace transformations
follows from the theory of 
the three point systems. This means, that also the transformations of
the four point scheme can be, in principle, derived from the three
point scheme. An open question is if the six point scheme and its
transformations
can be decomposed in a similar way.

Finally, we would like to mention possibility of considering (hierarchies of)
continuous deformations of the above lattice linear problems, which would 
lead to
(hierarchies of) discrete-differential integrable --- by construction ---
equations. Some aspects of deformations of the self-adjoint seven and five
linear systems have been elaborated in \cite{SND,Gilson,SDN-Toda}.

\section*{Acknowledgements}
The authors 
thank the Isaac Newton Institute for Mathematical Sciences
(Cambridge, UK) for hospitality during the programme 
\emph{Discrete Integrable Systems}.

\bibliographystyle{amsplain}

\providecommand{\bysame}{\leavevmode\hbox to3em{\hrulefill}\thinspace}

\end{document}